\documentclass{aastex}
\usepackage{graphics,graphicx}
\shorttitle{PAH in KPAIRs}
\shortauthors{Domingue et al.}

\begin{document}

 \title{Major Merger Galaxy Pairs at z=0: Dust Properties and Companion Morphology}  
 \author{Donovan L. Domingue\altaffilmark{1}, Chen Cao\altaffilmark{2,3}, C. Kevin Xu\altaffilmark{3},Thomas H. Jarrett\altaffilmark{4}, Joseph Ronca\altaffilmark{1}, Emily Hill\altaffilmark{1,5} and Allison Jacques\altaffilmark{1}}
%\email{donovan.domingue@gcsu.edu}

\altaffiltext{1}{Georgia College $\&$ State University, CBX 82, Milledgeville, GA 31061, USA}
\altaffiltext{2}{School of Space Science and Physics,
Shandong University, Weihai, Weihai, Shandong 264209, China}
\altaffiltext{3}{Infrared Processing and Analysis Center,
California Institute of Technology 100-22, Pasadena, CA 91125, USA}
\altaffiltext{4}{University of Cape Town, Private Bag X3, Rondebosch 7701, Republic of South Africa}
\altaffiltext{5}{Specialty Analytical, 1711 SE Capps Rd, Clackamas, OR  97015-6914, USA}

 \begin{abstract} 

We present an analysis of dust properties of a sample of close major-merger galaxy pairs selected by K$_{s}$ magnitude and redshift. The pairs represent the two populations of spiral-spiral (S$+$S) and mixed morphology spiral-elliptical (S$+$E). The CIGALE (Code Investigating GALaxy Emission) is used to fit dust models to the 2MASS, WISE and Herschel flux density measurements and derive the parameters describing the PAH contribution, interstellar radiation field (ISRF) and photo-dissociation regions (PDRs). Model fits verify our previous Spitzer Space Telescope analysis that S+S and S+E pairs do not have the same level of enhancement of star formation and differ in dust composition.  The spirals of mixed morphology galaxy pairs do not exhibit the enhancements in interstellar radiation field and therefore dust temperature for spirals in S$+$S pairs in contrast to what would be expected according to standard models of gas redistribution due to encounter torques. This suggests the importance of the companion environment/morphology in determining the dust properties of a spiral galaxy in a close major-merger pair. 

\end{abstract}  


 \section{Introduction}

 The Interstellar Medium (ISM) of spiral galaxies is composed of dust and gas at a multitude of temperatures as a result of the stellar life cycle. The various temperature states of the ISM along with HII regions (ionized hydrogen) provide an essential connection for understanding the evolution of galaxies due to their interaction with the radiation from young massive stars.  Graphites and silicates from 0.01 to 0.2$\mu$m in size compose the diffuse medium or cirrus which surrounds the star forming HII regions \citep{Mathis:1977, Draine:1984, Odonnell:1997}. While large dust grains obtain thermal equilibrium and exhibit a peak emission at wavelengths $\sim$100-300$\mu$m, smaller grains such as polycyclic aromatic hydrocarbons (PAH) grains are transiently heated by single photons and emit into the mid-infrared $\sim$3-30 $\mu$m wavelengths with particularly strong emission features in the 6-12 $\mu$m wavelength range \citep{Helou:2000, Draine:2007ta}. The large grain dust mixture is heated by both the interstellar radiation field (ISRF) leading to cirrus temperatures $\sim$20 K as well as the light from the star forming regions yielding a higher temperature dust component \citep{XuHelou, Dale2012}. Creation of the higher temperature dust component occurs at interface of the densest ISM component, the molecular clouds, and the ionized HII regions known as the photodissociation region (PDRs). The contribution of the PDR to the dust temperature is shown to be larger than that of previous studies after inclusion of Herschel Space Telescope \citep{Herschel} data in the dust temperature analysis of KINGFISH survey galaxies \citep{Dale2012}.
 Studies of star forming regions such as the Small Magellanic Cloud show a dependance of the PAH feature strengths on the ISRF intensity \citep{Contursi:2000}.
 Galaxies with increased specific star formation rates may destroy PAH molecules \citep{cook:2014, madden:2006, Engelbracht:2005}. Other causes of PAH destruction or absence have been attributed to metallicity effects \citep{Engelbracht:2005} and the presence of AGN \citep{Roche:AGN} while \citet{Alonso:AGNPAH} observe that PAH grains were not destroyed in six AGN systems in which the grains are possibly protected by high gas column densities in galaxy nuclei.

An important step in the process of potential evolution of galaxy ISM is the pairing of galaxies. Galaxy pairs are an initial step in the eventual merger of the two galaxies. The universe is composed of galaxies that have been undergoing this assembly process over the Hubble time \citep{Kauffmann:1993, Cole:2001}. The result is an effective change in the galaxy count with an ever-increasing distribution to larger mass representatives of galaxies \citep{Bundy:2004,Bundy:2009}. Star formation rates (SFR) and therefore the ISM evolve during mergers \citep{Brinchmann:1998,XuShupe:2012} as well as when the pair companions are in a pre-merger stage of interaction with SFR exhibiting an anti-correlation with close proximity \citep{Xu:1991,Barton:2000,Barton:2007,Scudder:2012}. The gravitational influence of galaxy interactions was first theorized by \citet{ToomreToomre:1972} and observationally shown in \citet{Larson:1977}. These processes may occur as gas is redistributed both in location and ISM phase by the gravitational interaction that takes place during the encounter \citep{Kennicutt:1987,Dasyra:2006,Hopkins:2006}. The standard model for gas redistribution is of a tidal torque delivering gas to the galaxy nucleus as in the simulations of \citet{Di-Matteo:2008}. Merging galaxies have increased velocity dispersion in their ISM \citep{Elmegreen:1995}. The increase in dispersion is dominated by supersonic gas turbulence in the cold star forming phase of the ISM \citep{Burkert:2006}. Turbulence and substructure induced in the ISM during galaxy mergers (see \citet{Bournaud:2011} for a review) is now an additional mechanism for star formation investigated through simulations \citep{Renaud:2014}.

The KPAIR sample of galaxy merger candidates was created in \citet{Domingue:2009mq} to establish a baseline of merger stage properties solely based on the galaxies physical proximity to each other. A key difference being that our sample will contain pairs in early and late stages of the galaxy-galaxy interaction while morphologically selected samples \citep{Conselice:2003,Darg2010} will be biased to actively distorted and changed systems. The choice of the galaxy sample location as the ``local" nearby universe lets us resolve the pairs into separate galaxies. Studies of distant systems are not always capable of making this distinction and our study can serve as a baseline for high redshift population studies.
Two major types of galaxies, spirals(S) and ellipticals (E), are represented in the pair sample. Three possible pair combinations are S$+$S, S$+$E and E$+$E. We have omitted the E$+$E option from further study as they do not represent a likely candidate for star formation enhancements due to lower relative ISM abundance. The KPAIR sample has been refined in an effort to understand the star forming properties in the observing campaigns of \citet{Xu:2010kpairs} and \citet{cao:2016} resulting in the current H-KPAIR sample examined in this paper.

\citet{cao:2016} demonstrate the spirals in S$+$S pairs show significant enhancement in sSFR and star formation efficiency (SFE) while spirals in S$+$E pairs do not exhibit these enhancements. \citet{XuShupe:2012} also found no significant sSFR enhancement in massive S$+$E pairs at any redshift. Other authors have shown this absence of SFR enhancements in S$+$E pairs and low SFR of spirals with early type neighbors \citep{Park:2008, Park:2009, Hwang:2011qw, Moon:2015}. \citet{Park:2008} suggest the hot x-ray halo of early type companions interacts hydrodynamically and deprives their paired spirals of cold gas and reduces SFRs. Another possible scenario for the relative lack of SFE enhancement in spirals of S$+$E pairs is the role of intrinsic interaction differences resulting from a disc-disc collision/ISM turbulence versus the single disc encounter. The likely unique merger history of a mixed-morphology pair (i.e. past major merger resulting in the early type) may include both of these scenarios. The available data from our IR studies allow us to further probe the properties of the dust component in the ISM of each population.

 We describe the sample characteristics chosen to examine the dust properties in these close major merger pairs populations and their relation to the star formation properties presented in \citet{cao:2016}. \citet{cao:2016} present the Far-Infrared (FIR) data used in the present analysis but concentrate on the derivation of SFR and dust mass as an indication of gas content with the fitting models of \citet{Draine:2007ta} (DL07). In this present work we take the further steps of adding the near and mid-infrared data to the analysis in order to understand the PAH contribution fractions and dust characteristics such as incident radiation intensity for both the pair and control samples. To describe our conclusions we first present the observations and data analysis required to create the flux density input to the SED models.Finally we present the best fit models and an analysis of their parameter correlations with the previously derived physical properties of the corresponding galaxies.

\section{Sample Selection}
The galaxy pair sample was created from matching the SDSS DR5 \citep{Adelman-McCarthy:2007} spectroscopic galaxy catalog with the Two Micron All Sky Survey (2MASS) Extended Source Catalog (XSC; \citet{Jarrett:2000}) as described in \citet{Domingue:2009mq}. The matched catalog created a list of candidate pairs for which redshift ($\Delta$v $<$ 1000 km s$^{-1}$), separation r, (5 h$^{-1}$ kpc $\leq$ r $\leq$ 20 h$^{-1}$ kpc) and mass ($\Delta$K$_{s}$ $<$ 1; mass ratio $\leq$ 2.5) are restricted to limit the sample to physical pairs complete to K$_{s}$ = 12.5. Spitzer Space telescope observations and analysis of a subsample of this final set of 170 major merger pairs is presented in \citet{Xu:2010kpairs}. Here as in \citet{cao:2016} we begin with the Herschel subset (H-KPAIRS) sample developed by further restricting the selection of pairs with three criteria : 1) keep only pairs where both pair members have spectroscopically confirmed redshifts to avoid false pairs, 2) remove elliptical+elliptical pairs to facilitate the study of star forming galaxies, and 3) keep only pairs with recession velocity v$>$2000 km s$^{-1}$. The resulting H-KPAIR sample contains 88 galaxy pairs (44 are Spiral +Spiral (S+S) and 44 are Spiral + Elliptical (S+E)) with a median redshift z$\sim$0.04.

The control galaxy sample was chosen from the HerMES survey \citep{Oliver:Hermes} for its Herschel coverage. It is selected based on matching morphology and stellar mass to the members of the H-KPAIR sample \citep{cao:2016}. HerMES (Bootes, EGS, ELAIS N1, Lockman SWIRE) field galaxies with companions at a projected distance of $<$ 70 kpc and $\Delta$M$_{star}$ $<$ 0.4 dex were rejected along with peculiar and low coverage (image edge) galaxies as an initial generation of the parent candidates . Spiral galaxies (morphology as in \citet{cao:2016}) were matched to pair members when $\Delta$M$_{star}$ $<$ 0.1 dex. Although the control galaxies were not selected to match in redshift, the final one-to-one choice of the closest redshift galaxy to the pair member allows for the final control sample to have a mean redshift within the standard deviation of the H-KPAIR sample. The H-KPAIR mean z=0.037$\pm$0.013 while that of the control is z=0.049$\pm$0.014. A similar comparison of the stellar mass confirms the mass match with a mean log(M$_{star}$ )=10.67$\pm$0.33 and 10.66$\pm$0.34 for the pair and control samples respectively.

\section{Data for Pairs}

The implementation of the DL07 models requires a range of photometric measurements to develop an SED from the near- to mid-infrared for our galaxy samples. This section describes the data gathered from the surveys including The Two Micron All Sky Survey (2MASS) \citep{2mass:mission} and the Wide Field Infrared Survey Explorer (WISE: \citet{Wise:mission}) as well as the Herschel Space Observatory (proposal ID: OT2\_cxu\_2) during which the pairs were observed in all six photometry bands from both Herschel-PACS \citep{PACS:2010} and Herschel-SPIRE \citep{SPIRE:2010} instruments.

\subsection{2MASS Data}
2MASS \citep{2mass:mission} observed the sky in the near-infrared at $J$ (1.25 $\mu$m), $H$ (1.65 $\mu$m) and $K_{s}$ (2.16 $\mu$m) bands which led to the development of the 2MASS  XSC. The original selection catalog of the sample pairs was developed with the magnitudes derived from the XSC while the pairs with separations of less than 30$\arcsec$ required the deblending techniques of profile fitting and subtraction applied in \citet{Domingue:2009mq}.  Both the catalog and profile fit magnitudes use the $K_{20}$ value for the $K_{s}$-band magnitude \citep{Jarrett:2000}. The resulting $K_{s}$-band photometry is included in our SED fits as a baseline for the stellar contribution to the mid-infrared.

\subsection{WISE Photometry and Calibrations}
Mid-Infrared catalog photometry and images were extracted from the WISE mission archives from images of the sky at four infrared wavelengths (3.4 $\mu$m, 4.6 $\mu$m, 12 $\mu$m, and 22 $\mu$m). Reprocessing the WISE ``Atlas" imaging to achieve improved resolution was a necessary step to reduce the blending of the small separation galaxy pair members. The WISE Imaging Point-Source FWHM is reduced from 11$\farcs$4 and 18$\farcs$6  to 3\farcs5 and 5\farcs5 at 12 $\mu$m(W3) and 22 $\mu$m(W4) respectively using the MCM-HiRes \citep{Masci:2009} techniques of \citet{jarrett:2012}.

WISE photometry on the H-KPAIRS sample was done as either standard aperture photometry or model fitting due to galaxy pair member overlap and blending. Aperture photometry was performed with APT (Aperture Photometry Tool, \cite{laher:2012}) when pairs were widely separated compared to the extended flux of the pair members. Annuli larger than each aperture were 
used for sky background determination and subtraction. Pixel masking within APT was incorporated when the companion galaxy occupied part of the appropriate sized annuli. Zero-point flux calibration and modifications to the error estimates are based on the guidelines in the Explanatory Supplement to the WISE Preliminary Data Release Products\footnote{http://wise2.ipac.caltech.edu/docs/release/prelim/expsup/wise\_prelrel\_toc.html}. When aperture photometry of the individual galaxies was not possible due to blending of the galaxies in a pair, aperture photometry was first applied to the entire pair and model fitting photometry was performed with IMFIT \citep{IMFIT} in order to retrieve the relative deblended flux density of each galaxy. Moffat model fits were adequate in matching the galaxy profiles in the WISE bands. The relative flux density of each galaxy is fairly robust to the parameter adjustments. Total aperture flux density of the pair was divided based on relative IMFIT results. Errors are based on the area dependent background error estimation.  Aperture corrections for extended sources of 0.97 and 1.03 are applied to the W3 and W4 bands respectively as recommended in \citet{jarrett:2013}. Color corrections are not applied to the flux density determination as the CIGALE code \citep{Noll:2009, Ciesla:2014} via its python implementation \citep{PCIGALE} incorporates the WISE filter transmission curves as well as the transmission curve of each band contributing to the measured SED. Photometric errors are the quadratic sum of background subtraction error and the rms error as calculated in \citet{Dale2012} with a modification to include the correlated noise. This noise-variance correction factor (F$_{corr}$) has a dependance on aperture radius specific to the W3 and W4 bands. The calibrations are based on the larger co-add pixels and conversions were made to HiRes pixels in the determination of the appropriate F$_{corr}$ for each aperture. With an assumption that background error is random, the modified error is,
\begin{equation}
\sigma = \sigma_{sky}(F_{corr})^{\frac{1}{2}}(N_{ap}+\frac{N_{ap}^2 }{N_{bk}})^{\frac{1}{2}}, 
\end{equation}
where $\sigma_{sky}$ is the background uncertainty and N$_{ap}$, N$_{bk}$ are the number of pixels in the aperture and background annulus respectively.
 
\begin{figure}[!htb]
\includegraphics[width=\textwidth,angle=0]{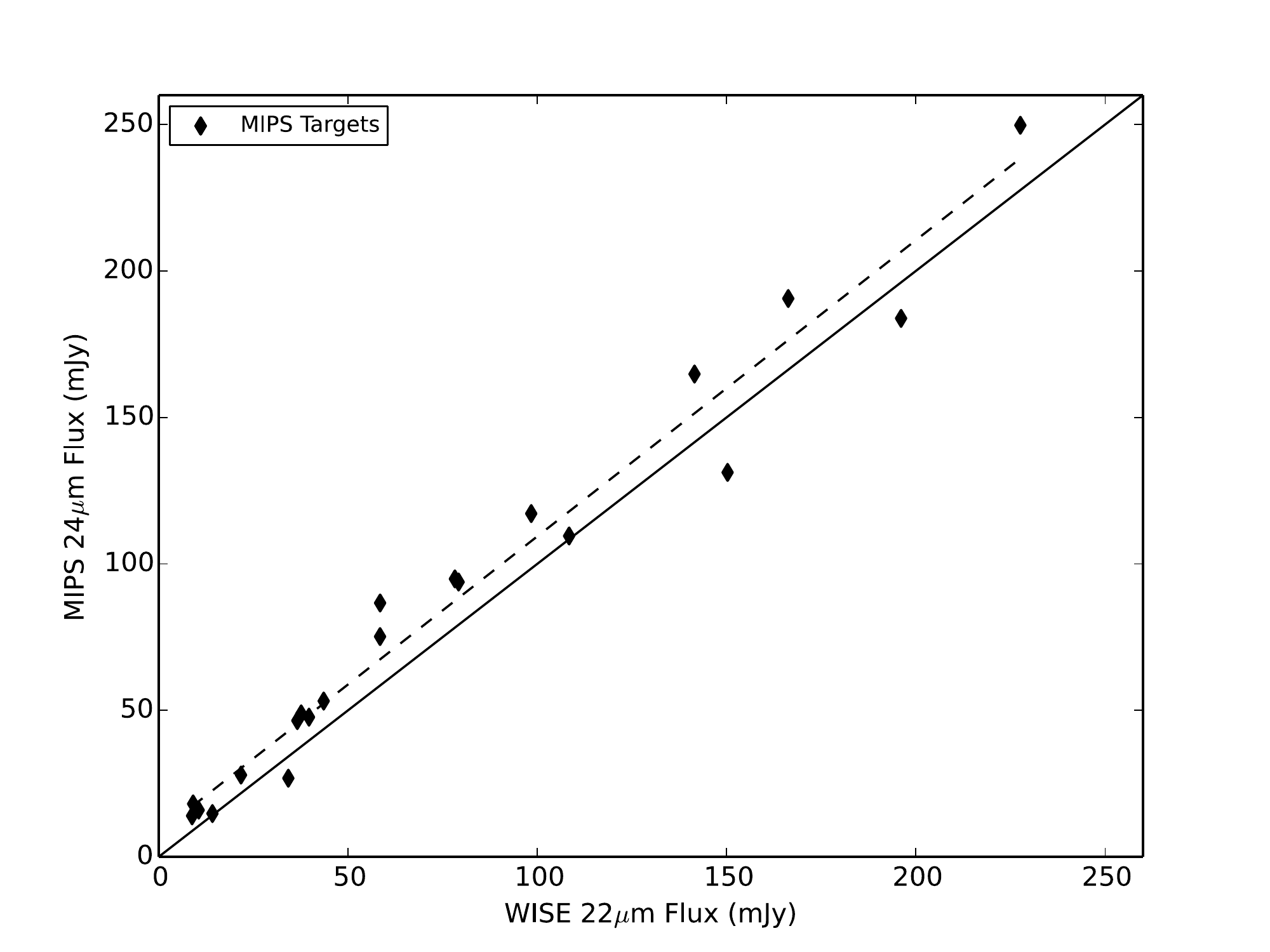}
\caption{The comparison of MIPS 24$\mu$m and WISE W4 22$\mu$m photometry for 22 targets in common with \citet{Xu:2010kpairs} The solid line represents a 1-to-1 match of the flux density and the dashed line is the best fit to the data. \label{fig:MIPS}
}
\end{figure}

 As a check on the validity of the W4 (22 $\mu$m) photometry, we compare the measured flux density to that of the Spitzer MIPS \citep{Rieke:2004} photometry for the pairs which were also a part of the Spitzer KPAIR study \citep{Xu:2010kpairs}. Figure \ref{fig:MIPS} displays the result of the comparison. Considering the 2$\mu$m wavelength difference of the flux density measures, the W4 values are consistent in a comparison to the MIPS flux densities \citep{Xu:2010kpairs}.
 
\subsection{Herschel Observations}
Photometry for the galaxy pairs was performed \citep{cao:2016} in six photometry bands from both Herschel-PACS (70,100,160 $\mu$m) \citep{PACS:2010} and Herschel-SPIRE (250, 350, 500$\mu$m) \citep{SPIRE:2010} instruments. Herschel PACS data were reduced with the application of UNIMAP \citep{Traficante:2011} to the HIPE \citep{Ott:2010} archived data. PACS map pixel sizes are 3$\farcs$2 at 70 and 100$\mu$m and 6$\farcs$4 at 160$\mu$m. Herschel SPIRE observations were reduced through HIPE 10.0.0 de-striper with standard SPIRE pipelines. SPIRE pixel sizes are 6\arcsec, 10\arcsec, 14\arcsec with beam FWHM of 18$\farcs$2, 24$\farcs$9, and 36$\farcs$3 for the 250, 350, 500$\mu$m images, respectively.

PACS and SPIRE photometry were performed through the use of aperture and background annulus flux density measures with APT \cite{laher:2012} or IDL code when pairs did not exhibit blending. Photometric errors are taken as a quadratic sum of the background subtraction error and rms error.
Pairs which were blended had a two step photometry procedure with the initial use of IMFIT\citep{IMFIT} to complete a simultaneous two component fit to the galaxy profiles. Exponential disks and gaussian fits were adequate to minimize model subtractions in PACS data while SPIRE data were reduced with PSF or 2-D gaussian models. In the PACS procedures as in the WISE photometry, large apertures on the entire pair are used to determine the combined flux density while the IMFIT models determine the relative contribution of each pair member. 

\section{Data for Control Sample}

\begin{figure}[!htb]
\includegraphics[width=\textwidth,angle=0]{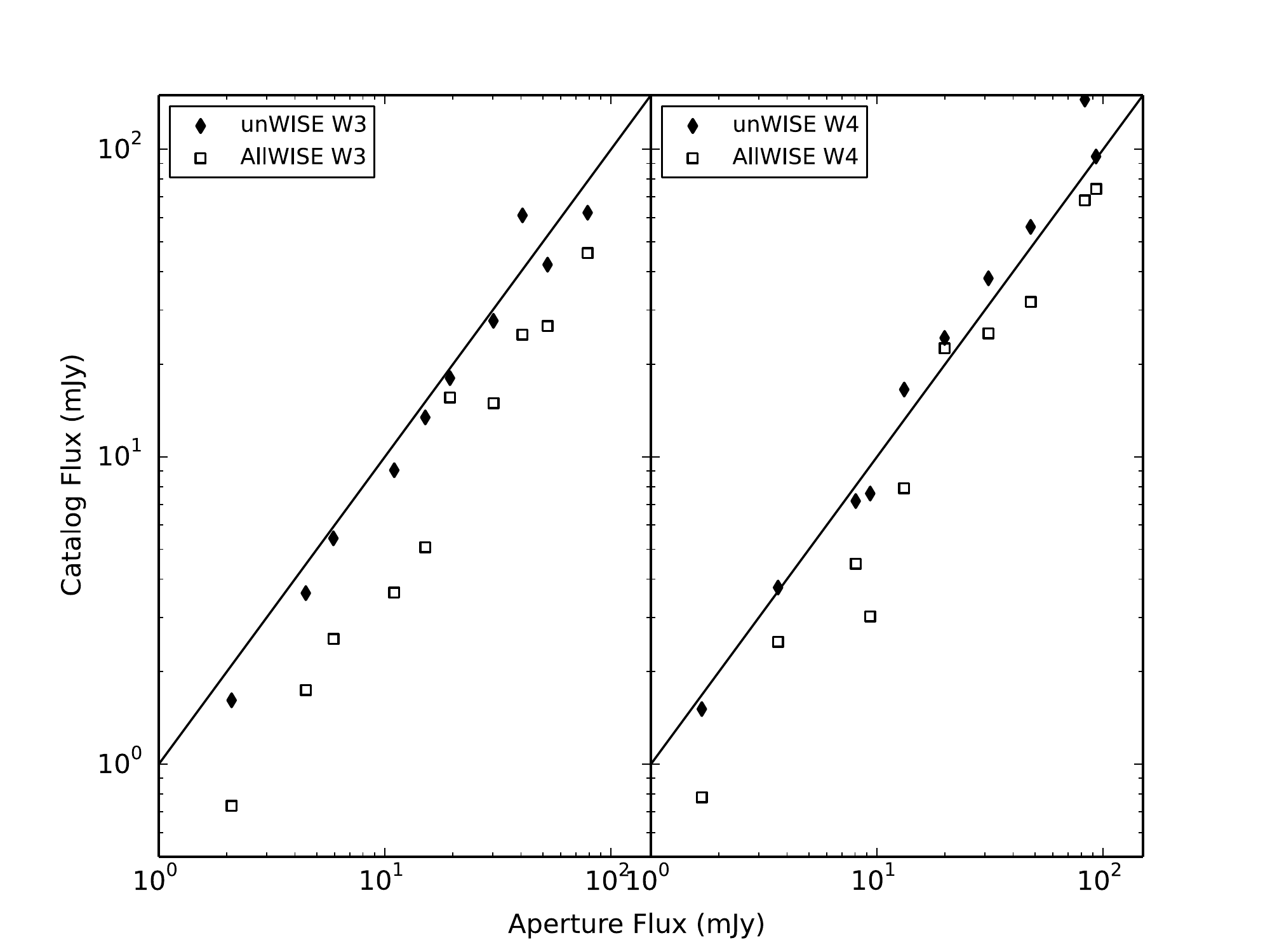}
\caption{(Left) WISE W3 flux densities from unWISE and allWISE catalogs compared to aperture flux density measures in this work. (Right) The same comparison for WISE W4 flux density measurements. The solid lines represent a 1-to-1 match of the flux density.\label{fig:WISEcheck}
}
\end{figure}

The lack of nearby neighbors for control sample galaxies simplifies the determination of their flux density in the same NIR-FIR bands used for the pair sample. Without the hindrance of pair blending, $K_{s}$ magnitudes are taken from the 2MASS XSC. The AllWISE \citep{AllWISE} source catalog provides magnitudes for W3 and W4 bands, however the catalog is optimized for point source detection and allows for active deblending which may split extended sources into two detections. An alternative AllWISE catalog magnitude for extended sources is the use of the gmagW* intended to use elliptical apertures derived from available 2MASS photometry. These magnitudes have been shown to underestimate the flux density in \citet{GAMA:WISE} while the choice of a single large aperture magnitude may not be appropriate for the varying size of the control galaxies in the W3 and W4 bands. Another source for catalog magnitudes derived from WISE data is the unWISE database of \citet{unwise:data, unwise:images} which uses forced photometry on objects identified in the SDSS-III Data Release 10 \citep{SDSS, SDSSIII}. We conduct a test comparison of WISE aperture flux density as conducted for the pair sample against unWISE and AllWISE flux densities for 10 of our control galaxies chosen to span their flux density range logarithmically. The unWISE flux densities provide the best match to the aperture flux densities as seen in Fig. \ref{fig:WISEcheck}. We have adopted the unWISE catalog flux densities for all control galaxies in the remaining CIGALE analysis.

Herschel data for the control sample are taken from the HerMES data release (v2) with 2\arcsec, 3\arcsec /pixel for PACS 100, 160$\mu$m data and 6\arcsec, 8$\farcs$33, and 12\arcsec /pixel  for SPIRE 250, 350, 500$\mu$m, respectively. Photometry was derived on these images \citep{cao:2016}with circular or elliptical apertures using IDL/phot. Background and photometric errors were determined from the aperture annuli. 

\section{Sample Characteristics}

We limit our candidates for SED analysis to the 132 spiral galaxies as representatives of a dust rich star forming population. In order to introduce available observation band uniformity to the SED analysis, we require that the spiral galaxies have PACS and SPIRE detections from 100$\mu$m-350$\mu$m as well as WISE 22$\mu$m and K$_{s}$ detections. Both upper limits and detections are initially used as input in the additional 12$\mu$m, 70 and 500$\mu$m bands for the galaxies meeting this criteria. The 70$\mu$m data is not available for our control galaxies and subsequent flux density input only include Herschel data from 100$\mu$m-500$\mu$m as discussed in section on SED fits. Based on the restriction, 47 spiral galaxies of the H-KPAIR sample are not included in our analysis. Of these excluded galaxies, 17 have been established to have log(sSFR) $<$ -11.5 \citep{cao:2016} with the implication that they are near the ``red and dead'' description of spirals. Another 15 only have an upper limit on SFR and the remaining 15 have missing flux density measures at Herschel bands determined as necessary for accurate dust property measurements.

\section{Dust Models and SED Fits} 
We use the python implementation of the SED fitting software known as the Code Investigating GALaxy Emission (CIGALE) \citep{PCIGALE, Roehlly:2011de, Noll:2009} to derive the PAH emission characteristics of the Herschel observed star forming galaxies in a $K_{s}$-band selected galaxy pair sample (H-KPAIRS) \cite{cao:2016}. The CIGALE code allows users to choose models for star formation history, single stellar populations and dust emission among other options to incorporate into a best-fit procedure for the observed SED. We implement the \citet{maraston:ssp} stellar population model with an initial mass function (IMF) of \citet{Kroupa:IMF} and a solar metallicity of $Z=0.02$ . The adopted star formation history model is composed of two decreasing exponentials with an e-folding time taken as a variable of 2 Gyr for "red and dead" spirals or 10 Gyr for active star forming spirals allowing for the best fit. The star forming history also allows a late burst mass fraction to vary between 0.001, 0.01, and 0.1. The dust model chosen here is the DL07 model as expanded in \citet{Aniano:2012} to include a larger range of models known as the DL2014 models in CIGALE. 
The varied parameters of the \citet{Draine:2007ta} models are qPAH, $\gamma$ ,and U$_{min}$. The parameter qPAH represents the percentage of the dust mass composed of polycyclic aromatic hydrocarbons (PAH) with $<$ 10$^{3}$ C atoms.  The (1-$\gamma$) is the fraction of dust exposed to the ISRF described by U$_{min}$, a dimensionless intensity factor. Other areas of the galaxy are exposed to an intensity factor U $\geq$ U$_{min}$ up to a maximum value of U$_{max}$ following a distribution of heating intensities from U$_{min}$ to U$_{max}$ described by the DL07 power law function;
 
 \begin{equation}
\frac{dM_{dust}}{dU} = (1-\gamma)M_{dust}\delta(U-U_{min})+\gamma M_{dust}\frac{\alpha-1}{U_{min}^{1-\alpha}-U_{max}^{1-\alpha}}U^{-\alpha},       
 \alpha \neq 1
\end{equation}
 where U$_{min}$ $\leq$ U $\leq$ U$_{max}$, $\delta$ is the Dirac delta function which ensures the first term only contributes outside of the PDR and M$_{dust}$ is the total dust mass. \citet{DraineDale:2007} indicate the power $\alpha$ may be held fixed $\alpha$ = 2 without affecting the quality of the fits to the SED of many galaxies and U$_{max}$ = 10$^{7}$ may also be set as a fixed parameter \citep{Aniano:2012}.
 The minimum ISRF (U$_{min}$) parameter is varied among all available values between U$_{min}$=0.1 to U$_{min}$=50. The qPAH model parameters are qPAH =0.47, 1.12, 1.77, 2.50, 3.19, 3.90, 4.58, 5.26, 5.95, 6.63, and 7.32. The $\gamma$ parameter as fraction of the galaxy is logarithmically spaced from 10$^{-3}$ to 1.0. The redshift of each galaxy is applied to create rest frame SED fits. 
 
 Two other quantities descriptive of the dust conditions \citep{Draine:2007ta} are the f$_{PDR}$, total dust luminosity that is emitted by the dust grains in regions with U$>$10$^{2}$, and the mean starlight intensity $\langle $U$\rangle$. The $\langle $U$\rangle$ and f$_{PDR}$ are defined as:
 \begin{equation}
\langle U \rangle = (1-\gamma)U_{min}+\frac{\gamma U_{min}ln(U_{max}/U_{min})}{1-U_{min}/U_{max}}
\end{equation}
and 
 \begin{equation}
f_{PDR} = \frac{\gamma ln(U_{max}/10^{2})}{(1-\gamma)(1-U_{min}/U_{max})+\gamma ln(U_{max}/U_{min})}.
\end{equation}

The CIGALE code utlitlizes $\chi^{2}$ minimization defined in \citet{Ciesla:2014} to find the best DL07 model match to the photometry of each galaxy as:
 \begin{equation}
\chi^{2}(a_{1}, ..., a_{i}, ...,a_{N})=\sum_{i}^{M}\left[\frac{y_{i}-\eta y(x_{i},a_{1}, ..., a_{i}, ...,a_{N})}{\sigma_{i}}\right]^{2},
\end{equation}
the reduced $\chi^{2}$ is:
\begin{equation}
\chi^{2}_{red}=\frac{\chi^{2}}{M-N}
\end{equation}
where the models are represented as y, the model parameters as a$_{i}$, the observations as x$_{i}$ and the observation errors are $\sigma_{i}$. There are N parameters and M number of data. The normalization $\eta$ is obtained from:
\begin{equation}
\eta=\frac{\sum_{i=1}^{N} y_{i} \times y(x_{i},a_{1}, ..., a_{i}, ...,a_{M})/\sigma_{i}^{2}}{\sum_{i=1}^{N} y(x_{i},a_{1}, ..., a_{i}, ...,a_{M})/\sigma_{i}^{2}}.
\end{equation}
A probability distribution function (PDF) is created as each parameter is varied and $\chi^{2}_{red}$ values are calculated. The minimum $\chi^{2}_{red}$ for each discrete parameter is used to develop the distribution from which the mean and standard deviation are taken as the ``estimated" value with its error. CIGALE reports the ``best" parameter as the parameter taken from the model with the minimum $\chi^{2}_{red}$. In this paper the ``estimated" values obtained from the PDFs are used for the dust analysis with their errors taken from the PDF standard deviations.

\begin{figure*}[!htb]
\includegraphics[width=\textwidth,angle=0]{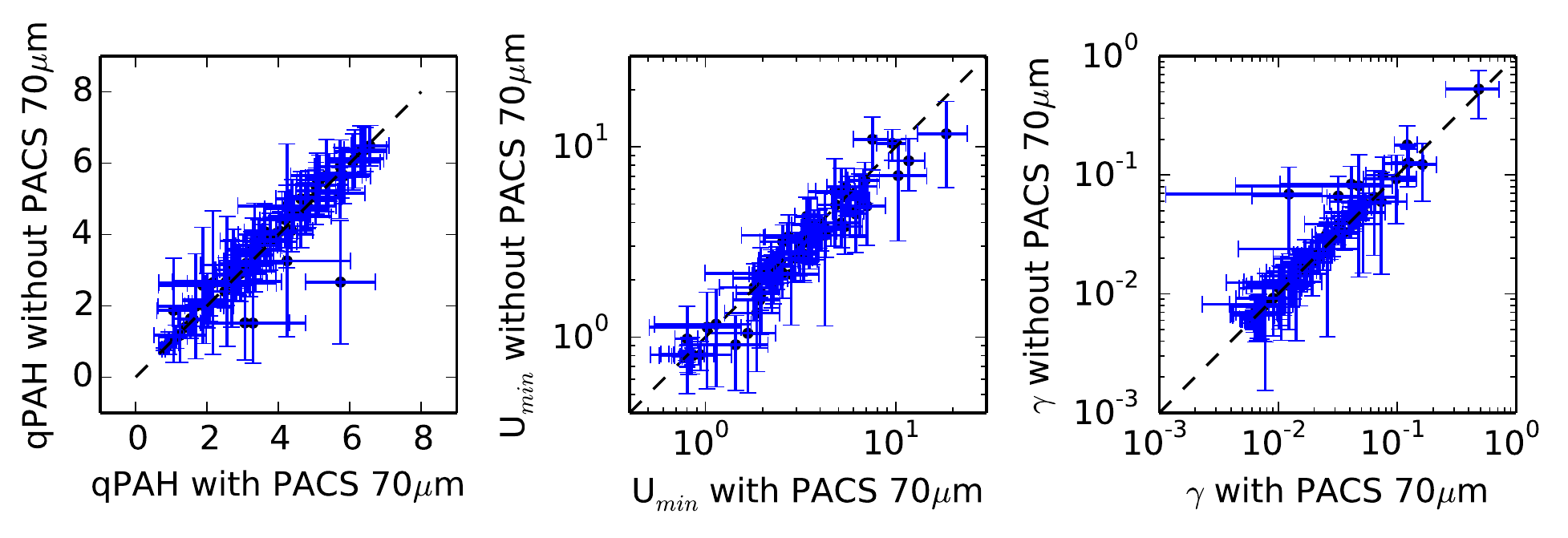}
\caption{A comparison of the parameter results for qPAH, U$_{min}$ and $\gamma$ from the use of CIGALE when including and excluding the available PACS 70 $\mu$m band as input. The dashed line represents a 1-to-1 match of the parameters. Error bars are from the probability distribution of each SED fit.\label{fig:compfits}
}
\end{figure*}

Since the control sample does not have the PACS 70$\mu$m flux density measurement as an input into the SED fit, we tested the derived parameters from our pair galaxy sample with and without the use of PACS 70$\mu$m flux densities. The parameter values for qPAH, U$_{min}$ and $\gamma$ derived from the full set of flux density measures are shown in Fig. \ref{fig:compfits} to be within the errors of these same parameters when the PACS 70$\mu$m is intentionally excluded from the fit with the exception of one galaxy fit. Due to the limits on the control sample, the CIGALE derived parameter analysis will be restricted to both samples without the PACS 70$\mu$m to increase the similarity of the precision on determination of the best models.

\begin{figure*}[!htb]
\includegraphics[width=\textwidth,angle=0]{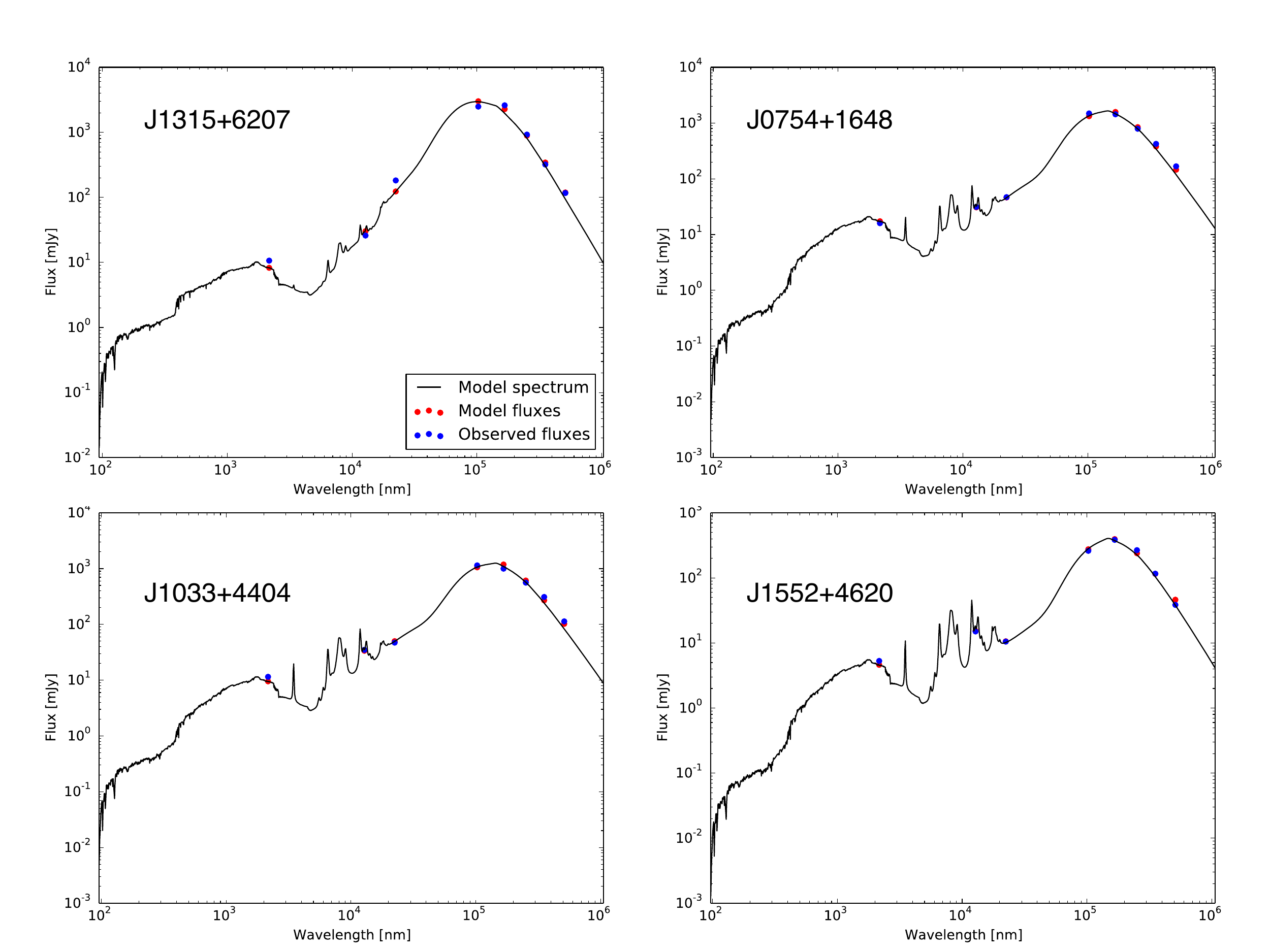}
\caption{Example SED fits for galaxies with a range of qPAH values exhibited by the intensity of the model spectrum emission lines.\label{fig:SEDfigure}
}
\end{figure*}

\citet{Ciesla:2014} demonstrate through the use of mock galaxy input, that missing 70$\mu$m data does not affect the determination of qPAH or $\gamma$ while U$_{min}$ is possibly overestimated by 18\%. Our analysis of U$_{min}$ treats each of our samples without the 70$\mu$m data and should present an accurate relative comparison of this dust parameter. The lack of the PAH 8$\mu$m feature as a measurement in our analysis will not affect any of the DL07 parameters \citep{Ciesla:2014}. Examples of the produced SED fits are shown in Fig. \ref{fig:SEDfigure}.

\section{Contributions of AGN}
As active galactic nuclei could contribute to the WISE bands and the overall SED fits and confuse the flux density contribution from dust, an analysis of the WISE flux densities as potential indicators of AGN activity was conducted. WISE photometry \citep{jarrett:2011,stern:2012, mateos:2012} has been used as a successful approach in AGN identification. A color criteria of Mateos et al. (2012), is applied using W1,W2 and W3 magnitudes taken from the unWISE forced photometry catalog \citep{unwise:data,unwise:images}. In the catalog, apertures are determined based on SDSS photometry and are held constant for the WISE band measures. The forced photometry should be sufficient to measure the magnitude differences W1-W2 and W2-W3 for the placement of the H-KPAIRs in the WISE color-color diagram.  From the full set of spirals in the H-KPAIRs, only J13151726+4424255 falls within the color-color diagram area associated with AGN in \citet{mateos:2012} and therefore the unWISE photometry indicate this may be our only AGN candidate. As a check on the catalog photometry analysis, additional manual aperture photometry is also carried out using APT on the W1 and W2 WISE imaging. Photometry was derived from the largest apertures allowed on the spirals which did not extend to the blended or overlap regions of the pairs. While these apertures do not contain the total flux density of each galaxy in many cases, they are kept the same for both bands and certainly include the nuclear contribution to these mid-infrared flux densities. According to the W1-W2 criteria, two galaxies are identified as a potential AGN by means of the aperture photometry, J12115648+4039184 and as also identified in the preliminary test, J13151726+4424255. These two galaxies have the highest values of $<$U$>$ and $\gamma$ in the sample and can be visually identified as the outlier points in Fig.\ref{fig:gammaUmin}.  The U$_{min}$ value of J13151726+4424255 is not unusual in the context of the reported U$_{min}$ distribution. The galaxies occupy two different mass bins of our subsequent analysis and any bias to our conclusions due to their inclusion is minimized.

\section{Distribution of Dust Parameters}

\begin{figure}[!htb]
\includegraphics[width=\textwidth,angle=0]{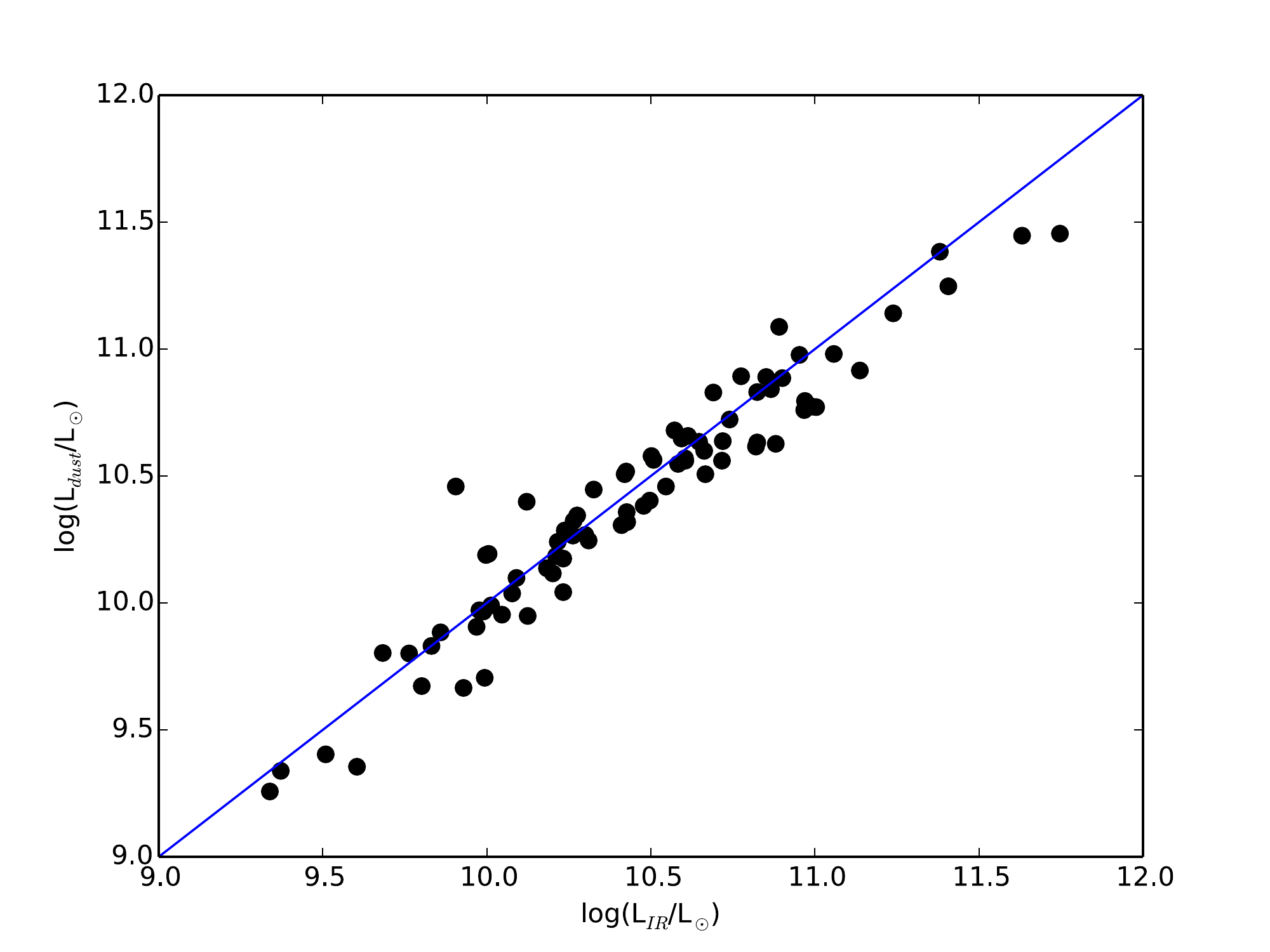}
\caption{A comparison of the CIGALE L$_{dust}$ to the Herschel band derived L$_{IR}$ of \citet{cao:2016}. The solid line represents a 1-to-1 match of the parameters.\label{fig:Lum_compare}
}
\end{figure}

\begin{figure*}[!htb]
\includegraphics[width=\textwidth,angle=0]{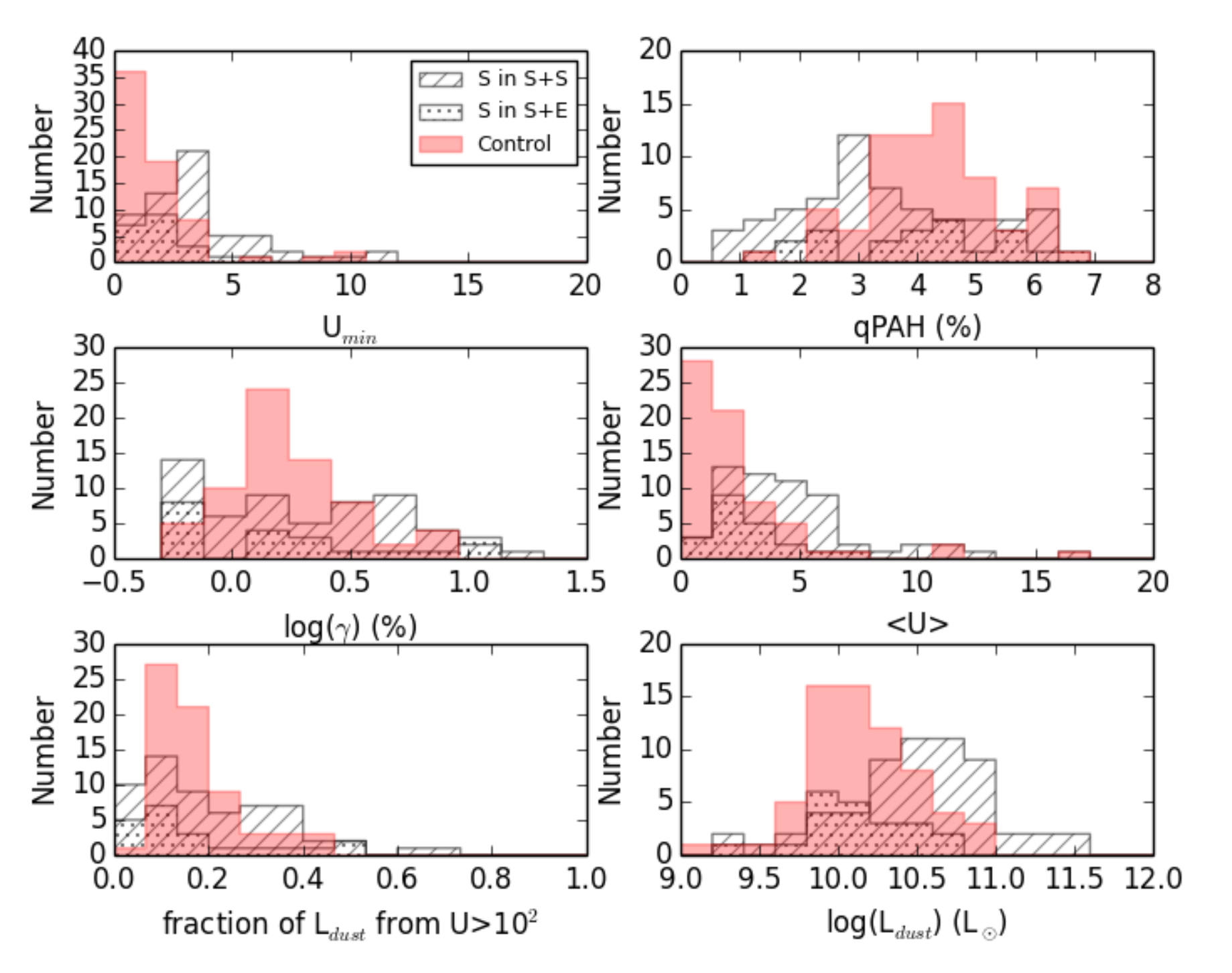}
\caption{(Histograms of the U$_{min}$, qPAH and $\gamma$ parameter output of CIGALE for the S$+$S, S$+$E and control samples along with those of the calculated values of $\langle$U$\rangle$, the fraction of L$_{dust}$ emitted from regions with U$>$10$^{2}$ (known as fPDR), and L$_{dust}$\label{fig:histograms}
}
\end{figure*}

The output of our CIGALE model fits is listed in Table 1. Of the 85 spirals input into our model fits, 5 can be classified as poorly fit with $\chi_{red}^{2}$$\geq$4. The average $\chi_{red}^{2}$$\sim$1 for the pair sample. The poorly fit galaxies are 3 from S$+$S and 2 from S$+$E categories. Analysis of the output parameters is limited to galaxies with $\chi_{red}^{2}$$<$4. We similarly reject analysis of control galaxies with $\chi_{red}^{2}$$\geq$4. This leaves 67 control galaxies for a remaining analysis. The CIGALE derived L$_{dust}$ is consistent (see Fig.\ref{fig:Lum_compare}) with the L$_{IR}$ of \citet{cao:2016}. Fig.\ref{fig:histograms} displays the distribution of the U$_{min}$, qPAH and $\gamma$ parameter output of CIGALE for the S$+$S, S$+$E and control samples along with those of the calculated values of $\langle$U$\rangle$, the fraction of L$_{dust}$ emitted from regions with U$>$10$^{2}$ (known as fPDR), and L$_{dust}$. A two-sided Kolmogorov-Smirnov (K-S) test (see Table 2) on the population distributions reveals those of the spirals in S$+$S pairs are statistically different from the control sample for basic output parameters of U$_{min}$, qPAH and L$_{dust}$. The distributions of the U$_{min}$ and L$_{dust}$ parameters for spirals in S$+$S are also different from their corresponding spirals in S$+$E. The qPAH distributions for the spirals in both pair morphology types differ at the P=0.15 significant level. The $\gamma$ parameter distributions show no significant differences across samples.

The calculated parameter distributions for $\langle$U$\rangle$ and fPDR also show a significant difference for S$+$S vs. the control sample while they are only different from S$+$E spirals in the $\langle$U$\rangle$ parameter distribution.

% table 1:
\begin{deluxetable}{ccccrcccccc}
\label{tbl:kpairdust}
%\tabletypesize{\normalsize}
%\tabletypesize{\tiny}
\tabletypesize{\scriptsize}
%\tiny(5pt);\scriptsize(7pt);\footnotesize(8pt);\small(9pt);\normalsize(10pt)
\setlength{\tabcolsep}{0.05in} %Tighten up the columns. See AASTeX FAQ
%\rotate
\tablenum{1}
\tablewidth{0pt}
%\tablewidth{0.95\linewidth}
\tablecaption{H-KPAIR Galaxy CIGALE SED Results}
\tablehead{
\colhead{(1)}   &\colhead{(2)}    &\colhead{(3)}    &\colhead{(4)}&\colhead{(5)}&\colhead{(6)} &\colhead{(7)}
             &\colhead{(8)}         &\colhead{(9)}   &\colhead{(10)}  &\colhead{(11)}  \\
\colhead{Galaxy ID}	&	\colhead{log(SFR/M$_{star}$)}	&	\colhead{log( M$_{star}$)}	&	\colhead{Type}	&	\colhead{$\chi$$_{red}^{2}$}	&	\colhead{qPAH}			&	\colhead{$\gamma$ }			&	\colhead{U$_{min}$}		&	\colhead{$<$U$>$}	&	\colhead{log(L$_{dust}$)}	&	\colhead{f$_{PDR}$}\\
\colhead{(2MASX)}	 & \colhead{yr$^{-1}$}	&\colhead{(M$\sun$)}	& & &\colhead{(\%)}	 &\colhead{(\%)}	 & & & \colhead{(L$\sun$)}& \\}
\startdata
J00202580+0049350	&	-10.00	&	10.70	&	S+E	&	0.17	&	3.47	$\pm$	0.55	&	1.21	$\pm$	0.51	&	3.31	$\pm$	0.61	&	3.87	&	10.10	&	0.12	\\
J01183417$-$0013416	&	-9.23	&	10.98	&	S+S	&	0.51	&	1.13	$\pm$	0.33	&	3.02	$\pm$	0.68	&	8.45	$\pm$	2.56	&	11.76	&	11.45	&	0.25	\\
J01183556$-$0013594	&	-9.88	&	10.65	&	S+S	&	0.98	&	3.43	$\pm$	0.66	&	0.82	$\pm$	0.66	&	3.80	$\pm$	0.86	&	4.22	&	10.72	&	0.08	\\
J02110638$-$0039191	&	-9.98	&	10.45	&	S+S	&	0.46	&	3.41	$\pm$	0.49	&	0.79	$\pm$	0.39	&	3.81	$\pm$	0.86	&	4.22	&	10.32	&	0.08	\\
J03381222+0110088	&	-10.42	&	11.05	&	S+E	&	0.41	&	4.41	$\pm$	0.78	&	1.21	$\pm$	0.55	&	2.18	$\pm$	0.57	&	2.55	&	10.51	&	0.12	\\
J07543194+1648214	&	-9.85	&	10.08	&	S+S	&	2.16	&	3.53	$\pm$	0.60	&	0.70	$\pm$	0.30	&	5.28	$\pm$	1.93	&	5.77	&	11.14	&	0.07	\\
J07543221+1648349	&	-9.98	&	10.34	&	S+S	&	0.57	&	2.86	$\pm$	0.55	&	1.13	$\pm$	0.59	&	2.60	$\pm$	0.66	&	3.02	&	10.98	&	0.11	\\
J08083377+3854534	&	-10.17	&	11.16	&	S+E	&	1.27	&	4.51	$\pm$	0.88	&	1.18	$\pm$	0.67	&	5.83	$\pm$	2.82	&	6.74	&	10.04	&	0.12	\\
J08233266+2120171	&	-10.27	&	10.56	&	S+S	&	0.28	&	3.69	$\pm$	0.72	&	3.41	$\pm$	0.83	&	3.61	$\pm$	0.77	&	5.32	&	10.14	&	0.27	\\
J08233421+2120515	&	-10.38	&	10.69	&	S+S	&	0.57	&	3.01	$\pm$	0.69	&	4.91	$\pm$	1.19	&	3.68	$\pm$	1.05	&	6.18	&	10.25	&	0.34	\\
J08291491+5531227	&	-9.96	&	11.12	&	S+S	&	1.08	&	5.37	$\pm$	0.65	&	0.60	$\pm$	0.03	&	1.05	$\pm$	0.54	&	1.14	&	10.29	&	0.06	\\
J08292083+5531081	&	-10.31	&	10.93	&	S+S	&	0.82	&	5.40	$\pm$	0.88	&	1.40	$\pm$	0.63	&	0.91	$\pm$	0.39	&	1.11	&	10.32	&	0.13	\\
J08381759+3054534	&	-10.44	&	10.49	&	S+S	&	2.02	&	3.18	$\pm$	0.79	&	2.10	$\pm$	0.63	&	6.17	$\pm$	1.56	&	7.89	&	10.51	&	0.19	\\
J08390125+3613042	&	-10.27	&	10.59	&	S+E	&	1.56	&	5.60	$\pm$	0.67	&	0.60	$\pm$	0.03	&	0.80	$\pm$	0.06	&	0.87	&	10.04	&	0.06	\\
J09060498+5144071	&	-10.14	&	10.42	&	S+E	&	0.34	&	5.53	$\pm$	0.70	&	0.60	$\pm$	0.07	&	0.81	$\pm$	0.14	&	0.88	&	9.95	&	0.06	\\
J09134606+4742001	&	-10.30	&	10.35	&	S+E	&	0.39	&	6.03	$\pm$	0.79	&	2.69	$\pm$	0.74	&	2.17	$\pm$	0.35	&	3.01	&	10.65	&	0.22	\\
J09155467+4419510	&	-10.95	&	10.84	&	S+S	&	1.75	&	1.99	$\pm$	0.56	&	5.38	$\pm$	1.22	&	6.20	$\pm$	1.46	&	10.63	&	11.25	&	0.36	\\
J09155552+4419580	&	-10.49	&	11.19	&	S+S	&	1.60	&	1.40	$\pm$	0.07	&	0.60	$\pm$	0.03	&	10.43	$\pm$	1.93	&	11.23	&	11.45	&	0.06	\\
J09374413+0245394	&	-10.63	&	10.88	&	S+E	&	0.17	&	3.86	$\pm$	0.61	&	1.29	$\pm$	0.52	&	2.35	$\pm$	0.46	&	2.78	&	10.76	&	0.13	\\
J10100079+5440198	&	-10.39	&	9.84	&	S+S	&	0.16	&	4.67	$\pm$	0.68	&	0.68	$\pm$	0.28	&	2.28	$\pm$	0.37	&	2.50	&	10.89	&	0.07	\\
J10100212+5440279	&	-10.64	&	10.22	&	S+S	&	0.83	&	3.81	$\pm$	1.06	&	1.27	$\pm$	0.87	&	1.34	$\pm$	0.68	&	1.59	&	10.38	&	0.12	\\
J10205188+4831096	&	-10.20	&	10.12	&	S+E	&	0.35	&	2.41	$\pm$	0.46	&	2.19	$\pm$	0.60	&	3.36	$\pm$	1.02	&	4.39	&	10.34	&	0.19	\\
J10225647+3446564	&	-10.07	&	10.06	&	S+S	&	0.23	&	1.64	$\pm$	0.55	&	5.52	$\pm$	1.24	&	2.74	$\pm$	0.75	&	4.87	&	10.64	&	0.36	\\
J10233658+4220477	&	-10.75	&	10.16	&	S+S	&	1.73	&	2.63	$\pm$	0.63	&	4.70	$\pm$	1.18	&	3.22	$\pm$	0.94	&	5.33	&	10.98	&	0.33	\\
J10233684+4221037	&	-10.30	&	10.91	&	S+S	&	0.75	&	5.65	$\pm$	1.01	&	1.24	$\pm$	0.84	&	4.89	$\pm$	1.86	&	5.71	&	10.46	&	0.12	\\
J10272950+0114490	&	-10.81	&	10.98	&	S+E	&	0.56	&	4.73	$\pm$	0.71	&	0.70	$\pm$	0.30	&	4.95	$\pm$	1.23	&	5.42	&	10.17	&	0.07	\\
J10332972+4404342	&	-10.57	&	10.60	&	S+S	&	1.17	&	3.69	$\pm$	0.58	&	1.69	$\pm$	0.55	&	3.18	$\pm$	0.77	&	3.92	&	10.92	&	0.16	\\
J10333162+4404212	&	-10.51	&	10.86	&	S+S	&	1.41	&	3.12	$\pm$	0.44	&	0.61	$\pm$	0.09	&	3.34	$\pm$	0.65	&	3.63	&	10.55	&	0.06	\\
J10435053+0645466	&	-10.19	&	10.50	&	S+S	&	1.17	&	3.01	$\pm$	0.71	&	5.69	$\pm$	1.36	&	2.82	$\pm$	0.73	&	5.07	&	10.66	&	0.36	\\
J10435268+0645256	&	-9.78	&	10.81	&	S+S	&	0.99	&	4.80	$\pm$	1.74	&	3.87	$\pm$	3.91	&	0.79	$\pm$	0.04	&	1.26	&	9.80	&	0.28	\\
J10452478+3910298	&	-10.47	&	10.14	&	S+E	&	3.40	&	1.87	$\pm$	1.47	&	12.26	$\pm$	6.23	&	0.79	$\pm$	0.04	&	2.29	&	9.80	&	0.49	\\
J11065068+4751090	&	-10.03	&	10.48	&	S+S	&	0.25	&	5.90	$\pm$	0.75	&	0.68	$\pm$	0.28	&	3.13	$\pm$	0.69	&	3.43	&	10.89	&	0.07	\\
J11204657+0028142	&	-10.12	&	10.38	&	S+S	&	0.22	&	2.14	$\pm$	0.76	&	9.95	$\pm$	4.22	&	7.06	$\pm$	3.85	&	16.30	&	9.71	&	0.50	\\
J11251716+0226488	&	-9.65	&	10.67	&	S+S	&	1.09	&	1.52	$\pm$	1.05	&	6.65	$\pm$	3.03	&	0.81	$\pm$	0.16	&	1.64	&	10.19	&	0.38	\\
J11273289+3604168	&	-11.05	&	11.26	&	S+S	&	3.34	&	1.52	$\pm$	1.12	&	8.37	$\pm$	3.47	&	0.82	$\pm$	0.18	&	1.86	&	9.83	&	0.42	\\
J11273467+3603470	&	-9.70	&	10.96	&	S+S	&	1.21	&	5.73	$\pm$	0.71	&	0.60	$\pm$	0.06	&	2.28	$\pm$	0.39	&	2.48	&	10.60	&	0.06	\\
J11440433+3332339	&	-10.37	&	10.92	&	S+E	&	0.75	&	4.80	$\pm$	0.64	&	0.62	$\pm$	0.15	&	2.31	$\pm$	0.57	&	2.51	&	9.97	&	0.07	\\
J11484370+3547002	&	-10.07	&	11.15	&	S+S	&	0.52	&	2.68	$\pm$	0.52	&	2.15	$\pm$	0.66	&	2.04	$\pm$	0.43	&	2.67	&	10.84	&	0.19	\\
J11484525+3547092	&	-10.28	&	10.74	&	S+S	&	1.76	&	2.65	$\pm$	2.01	&	5.99	$\pm$	4.52	&	2.17	$\pm$	0.78	&	4.03	&	10.83	&	0.37	\\
J11501399+3746306	&	-9.32	&	10.74	&	S+S	&	1.75	&	6.12	$\pm$	0.81	&	1.11	$\pm$	0.61	&	2.02	$\pm$	0.68	&	2.35	&	10.40	&	0.11	\\
J12020424+5342317	&	-10.35	&	10.49	&	S+E	&	0.35	&	1.99	$\pm$	1.47	&	9.29	$\pm$	5.41	&	0.80	$\pm$	0.11	&	1.94	&	10.12	&	0.44	\\
J12115507+4039182	&	-9.47	&	11.11	&	S+S	&	0.74	&	2.62	$\pm$	1.23	&	2.39	$\pm$	1.95	&	5.82	$\pm$	1.86	&	7.68	&	9.97	&	0.21	\\
J12115648+4039184	&	-10.08	&	10.89	&	S+S	&	1.51	&	0.95	$\pm$	0.32	&	17.89	$\pm$	8.14	&	11.71	$\pm$	5.62	&	38.23	&	10.63	&	0.63	\\
J12191866+1201054	&	-10.58	&	10.71	&	S+E	&	0.67	&	4.00	$\pm$	0.87	&	1.85	$\pm$	0.88	&	2.02	$\pm$	0.52	&	2.56	&	9.88	&	0.17	\\
J12433887+4405399	&	-10.63	&	10.91	&	S+E	&	1.59	&	4.03	$\pm$	0.50	&	0.60	$\pm$	0.03	&	1.70	$\pm$	0.67	&	1.85	&	9.99	&	0.06	\\
J12525011+4645272	&	-10.50	&	10.60	&	S+E	&	1.54	&	1.18	$\pm$	0.77	&	12.52	$\pm$	4.55	&	0.82	$\pm$	0.18	&	2.39	&	10.24	&	0.49	\\
J13011662+4803366	&	-10.32	&	10.84	&	S+S	&	1.14	&	3.58	$\pm$	0.87	&	4.52	$\pm$	1.33	&	3.98	$\pm$	1.02	&	6.45	&	10.56	&	0.32	\\
J13011835+4803304	&	-9.91	&	10.66	&	S+S	&	0.86	&	2.66	$\pm$	0.69	&	1.98	$\pm$	0.89	&	4.60	$\pm$	1.27	&	5.83	&	10.36	&	0.18	\\
J13082737+0422125	&	-9.83	&	10.81	&	S+S	&	0.03	&	3.26	$\pm$	2.12	&	8.11	$\pm$	6.72	&	0.80	$\pm$	0.06	&	1.79	&	9.34	&	0.42	\\
J13082964+0422045	&	-9.97	&	10.55	&	S+S	&	0.15	&	2.68	$\pm$	1.82	&	6.50	$\pm$	4.39	&	0.80	$\pm$	0.10	&	1.60	&	9.40	&	0.37	\\
J13151386+4424264	&	-9.88	&	10.53	&	S+S	&	0.79	&	6.36	$\pm$	0.67	&	0.82	$\pm$	0.42	&	2.38	$\pm$	0.60	&	2.66	&	10.45	&	0.08	\\
J13151726+4424255	&	-10.67	&	10.42	&	S+S	&	0.36	&	3.14	$\pm$	1.01	&	52.74	$\pm$	22.91	&	3.43	$\pm$	2.29	&	28.57	&	11.09	&	0.73	\\
J13153076+6207447	&	-10.18	&	11.11	&	S+S	&	2.56	&	0.78	$\pm$	0.04	&	1.57	$\pm$	0.33	&	10.95	$\pm$	3.38	&	13.13	&	10.89	&	0.15	\\
J13153506+6207287	&	-10.92	&	10.97	&	S+S	&	6.58	&	0.79	$\pm$	0.07	&	3.46	$\pm$	0.66	&	17.38	$\pm$	3.23	&	24.76	&	11.35	&	0.28	\\
J13325525$-$0301347	&	-10.05	&	10.94	&	S+S	&	0.99	&	2.06	$\pm$	0.49	&	3.91	$\pm$	0.91	&	3.81	$\pm$	1.10	&	5.87	&	10.62	&	0.29	\\
J13325655$-$0301395	&	-10.42	&	11.27	&	S+S	&	0.76	&	4.57	$\pm$	0.53	&	0.60	$\pm$	0.03	&	3.27	$\pm$	0.67	&	3.55	&	10.57	&	0.06	\\
J13462001$-$0325407	&	-10.80	&	10.63	&	S+E	&	4.37	&	6.59	$\pm$	0.54	&	0.64	$\pm$	0.19	&	1.03	$\pm$	0.52	&	1.13	&	9.51	&	0.07	\\
J14005783+4251203	&	-10.41	&	10.95	&	S+S	&	0.25	&	3.01	$\pm$	0.52	&	2.11	$\pm$	0.56	&	3.18	$\pm$	0.66	&	4.11	&	10.63	&	0.19	\\
J14005879+4250427	&	-10.37	&	10.40	&	S+S	&	0.91	&	2.07	$\pm$	0.50	&	4.32	$\pm$	0.96	&	5.63	$\pm$	1.35	&	8.89	&	10.80	&	0.32	\\
J14055079+6542598	&	-10.77	&	10.84	&	S+E	&	0.65	&	5.00	$\pm$	0.76	&	0.70	$\pm$	0.30	&	0.80	$\pm$	0.08	&	0.88	&	9.67	&	0.07	\\
J14062157+5043303	&	-9.97	&	10.15	&	S+E	&	0.23	&	5.64	$\pm$	0.77	&	0.66	$\pm$	0.24	&	3.77	$\pm$	0.65	&	4.12	&	10.46	&	0.07	\\
J14070703$-$0234513	&	-10.79	&	11.05	&	S+E	&	0.05	&	2.66	$\pm$	1.74	&	6.91	$\pm$	4.73	&	0.98	$\pm$	0.47	&	2.00	&	10.40	&	0.39	\\
J14234238+3400324	&	-9.51	&	10.36	&	S+S	&	0.35	&	5.00	$\pm$	0.79	&	1.16	$\pm$	0.55	&	3.22	$\pm$	0.74	&	3.74	&	9.67	&	0.11	\\
J14234632+3401012	&	-9.76	&	10.51	&	S+S	&	1.60	&	6.38	$\pm$	0.61	&	0.61	$\pm$	0.11	&	1.82	$\pm$	0.61	&	1.99	&	9.26	&	0.06	\\
J14245831$-$0303597	&	-9.84	&	10.31	&	S+S	&	1.01	&	5.16	$\pm$	0.71	&	0.60	$\pm$	0.06	&	4.31	$\pm$	1.13	&	4.66	&	10.68	&	0.06	\\
J14295031+3534122	&	-10.36	&	10.73	&	S+S	&	1.56	&	1.61	$\pm$	0.43	&	2.73	$\pm$	0.68	&	3.90	$\pm$	1.42	&	5.36	&	9.91	&	0.23	\\
J14334683+4004512	&	-10.12	&	11.04	&	S+S	&	0.53	&	4.08	$\pm$	0.68	&	1.44	$\pm$	0.53	&	1.95	$\pm$	0.52	&	2.35	&	10.56	&	0.14	\\
J14334840+4005392	&	-9.74	&	10.62	&	S+S	&	0.54	&	5.04	$\pm$	0.77	&	0.92	$\pm$	0.48	&	5.44	$\pm$	1.09	&	6.12	&	10.83	&	0.09	\\
\enddata
%\tablecomments{Description of Columns: (1)Galaxy ID taken from 2MASS. (2) log(sSFR) from \citet{cao:2016}. (3) Stellar Mass. (4) Pair Type:S+S or S+E. (5) reduced $\chi$$^{2}$ of model fit. (6) Percentage of dust in PAH form. (7) Percentage of galaxy dust exposed to U$>$U$_{min}$. (8) Relative intensity of the diffuse ISRF. (9) Average relative intensity ,U, of starlight on dust grains. (10) Luminosity of dust. (11) fraction of luminosity due to regions with U$>$$10^{2}$.}
\end{deluxetable}

\begin{deluxetable}{ccccrcccccc}
\label{tbl:kpairdust}
%\tabletypesize{\normalsize}
%\tabletypesize{\tiny}
\tabletypesize{\scriptsize}
%\tiny(5pt);\scriptsize(7pt);\footnotesize(8pt);\small(9pt);\normalsize(10pt)
\setlength{\tabcolsep}{0.05in} %Tighten up the columns. See AASTeX FAQ
%\rotate
\tablenum{1}
\tablewidth{0pt}
%\tablewidth{0.95\linewidth}
\tablecaption{(Continued)}
\tablehead{
\colhead{(1)}   &\colhead{(2)}    &\colhead{(3)}    &\colhead{(4)}&\colhead{(5)}&\colhead{(6)} &\colhead{(7)}
             &\colhead{(8)}         &\colhead{(9)}   &\colhead{(10)}  &\colhead{(11)}  \\
\colhead{Galaxy ID}	&	\colhead{log(SFR/M$_{star}$)}	&	\colhead{log( M$_{star}$)}	&	\colhead{Type}	&	\colhead{$\chi$$_{red}^{2}$}	&	\colhead{qPAH}			&	\colhead{$\gamma$ }			&	\colhead{U$_{min}$}		&	\colhead{$<$U$>$}	&	\colhead{log(L$_{dust}$)}	&	\colhead{f$_{PDR}$}\\
\colhead{(2MASX)}	 & \colhead{yr$^{-1}$}	&\colhead{(M$\sun$)}	& & &\colhead{(\%)}	 &\colhead{(\%)}	 & & & \colhead{(L$\sun$)}& \\}
\startdata

J14442055+1207429	&	-9.06	&	10.80	&	S+S	&	3.25	&	0.87	$\pm$	0.24	&	12.67	$\pm$	3.86	&	1.17	$\pm$	0.62	&	3.39	&	10.19	&	0.50	\\
J14442079+1207552	&	-9.77	&	10.62	&	S+S	&	1.02	&	2.96	$\pm$	0.70	&	3.30	$\pm$	0.98	&	2.57	$\pm$	0.63	&	3.77	&	10.58	&	0.26	\\
J15064579+0346214	&	-10.27	&	10.91	&	S+S	&	0.81	&	6.32	$\pm$	0.72	&	2.85	$\pm$	0.80	&	1.13	$\pm$	0.59	&	1.61	&	10.27	&	0.23	\\
J15101776+5810375	&	-9.80	&	10.71	&	S+S	&	0.36	&	4.11	$\pm$	0.60	&	0.67	$\pm$	0.25	&	3.17	$\pm$	0.72	&	3.46	&	10.27	&	0.07	\\
J15233768+3749030	&	-11.24	&	10.69	&	S+E	&	0.45	&	2.57	$\pm$	1.63	&	4.84	$\pm$	3.35	&	2.14	$\pm$	0.98	&	3.63	&	9.35	&	0.33	\\
J15281276+4255474	&	-9.61	&	10.60	&	S+S	&	0.48	&	4.09	$\pm$	0.56	&	0.60	$\pm$	0.05	&	2.54	$\pm$	0.61	&	2.76	&	10.52	&	0.06	\\
J15523393+4620237	&	-10.44	&	10.30	&	S+E	&	0.40	&	6.49	$\pm$	0.57	&	0.61	$\pm$	0.11	&	2.12	$\pm$	0.33	&	2.30	&	10.56	&	0.06	\\
J15562191+4757172	&	-10.18	&	10.14	&	S+E	&	0.21	&	3.49	$\pm$	0.60	&	2.14	$\pm$	0.60	&	2.57	$\pm$	0.63	&	3.35	&	9.95	&	0.19	\\
J15583784+3227471	&	-10.30	&	10.94	&	S+S	&	0.37	&	2.90	$\pm$	0.78	&	0.60	$\pm$	0.03	&	3.91	$\pm$	0.73	&	4.23	&	10.63	&	0.06	\\
J16024254+4111499	&	-10.07	&	10.79	&	S+S	&	1.25	&	4.72	$\pm$	0.83	&	1.69	$\pm$	0.65	&	3.32	$\pm$	0.77	&	4.09	&	10.77	&	0.16	\\
J16024475+4111589	&	-10.70	&	10.76	&	S+S	&	0.28	&	3.79	$\pm$	0.79	&	1.53	$\pm$	0.74	&	3.18	$\pm$	0.68	&	3.86	&	10.31	&	0.15	\\
J16080648+2529066	&	-10.57	&	11.11	&	S+S	&	13.90	&	6.63	$\pm$	0.54	&	2.88	$\pm$	0.77	&	0.80	$\pm$	0.15	&	1.15	&	10.02	&	0.23	\\
J16082261+2328459	&	-10.49	&	10.96	&	S+S	&	2.83	&	5.22	$\pm$	0.65	&	0.60	$\pm$	0.03	&	1.57	$\pm$	0.69	&	1.71	&	10.19	&	0.06	\\
J16082354+2328240	&	-10.21	&	10.89	&	S+S	&	1.88	&	2.18	$\pm$	0.46	&	3.16	$\pm$	0.84	&	4.44	$\pm$	1.65	&	6.36	&	10.77	&	0.25	\\
J16372583+4650161	&	-10.83	&	11.16	&	S+S	&	4.83	&	1.26	$\pm$	0.83	&	8.47	$\pm$	3.21	&	0.79	$\pm$	0.04	&	1.82	&	10.33	&	0.42	\\
J17045097+3449020	&	-9.60	&	10.99	&	S+S	&	0.49	&	3.19	$\pm$	0.82	&	4.17	$\pm$	1.22	&	6.68	$\pm$	1.62	&	10.35	&	11.38	&	0.31	\\
J20471908+0019150	&	-10.83	&	11.09	&	S+E	&	8.29	&	4.02	$\pm$	0.55	&	0.60	$\pm$	0.03	&	0.79	$\pm$	0.04	&	0.87	&	9.71	&	0.06	\\
\enddata
\tablecomments{Description of Columns: (1)Galaxy ID taken from 2MASS. (2) log(sSFR) from \citet{cao:2016}. (3) Stellar Mass. (4) Pair Type:S+S or S+E. (5) reduced $\chi$$^{2}$ of model fit. (6) Percentage of dust in PAH form. (7) Percentage of galaxy dust exposed to U$>$U$_{min}$. (8) Relative intensity of the diffuse ISRF. (9) Average relative intensity ,U, of starlight on dust grains. (10) Luminosity of dust. (11) fraction of luminosity due to regions with U$>$$10^{2}$.}
\end{deluxetable}

\begin{deluxetable}{lcccccc}
\label{tbl:stats}\tabletypesize{\scriptsize}
%\tiny(5pt);\scriptsize(7pt);\footnotesize(8pt);\small(9pt);\normalsize(10pt)
\setlength{\tabcolsep}{0.05in} %Tighten up the columns. See AASTeX FAQ
%\rotate
\tablenum{2}
\tablewidth{0pt}
\tablecaption{K-S test Distribution Statistics}
\tablehead{
&P$_{Umin}$&P$_{qPAH}$&P$_{\gamma}$&P$_{\langle U \rangle}$&P$_{fPDR}$&P$_{Ldust}$\\}
\startdata
S$+$S vs. Control&2.3x10$^{-7}$&5.5x10$^{-5}$&0.041&6.1x10$^{-7}$&0.023&1.6x10$^{-5}$\\
S$+$E vs. Control&0.081&0.522&0.079&0.010&0.079&0.997\\
S$+$S vs. S$+$E&0.005&0.146&0.614&0.012&0.614&0.002\\

\enddata
\tablecomments{The K-S test significance P for distribution similarity of each of the histograms in Fig.\ref{fig:histograms} versus the other samples as described in column 1.}
\end{deluxetable}

\section{Correlation of Dust Parameters}
\begin{figure}[!htb]
\includegraphics[width=\textwidth,angle=0]{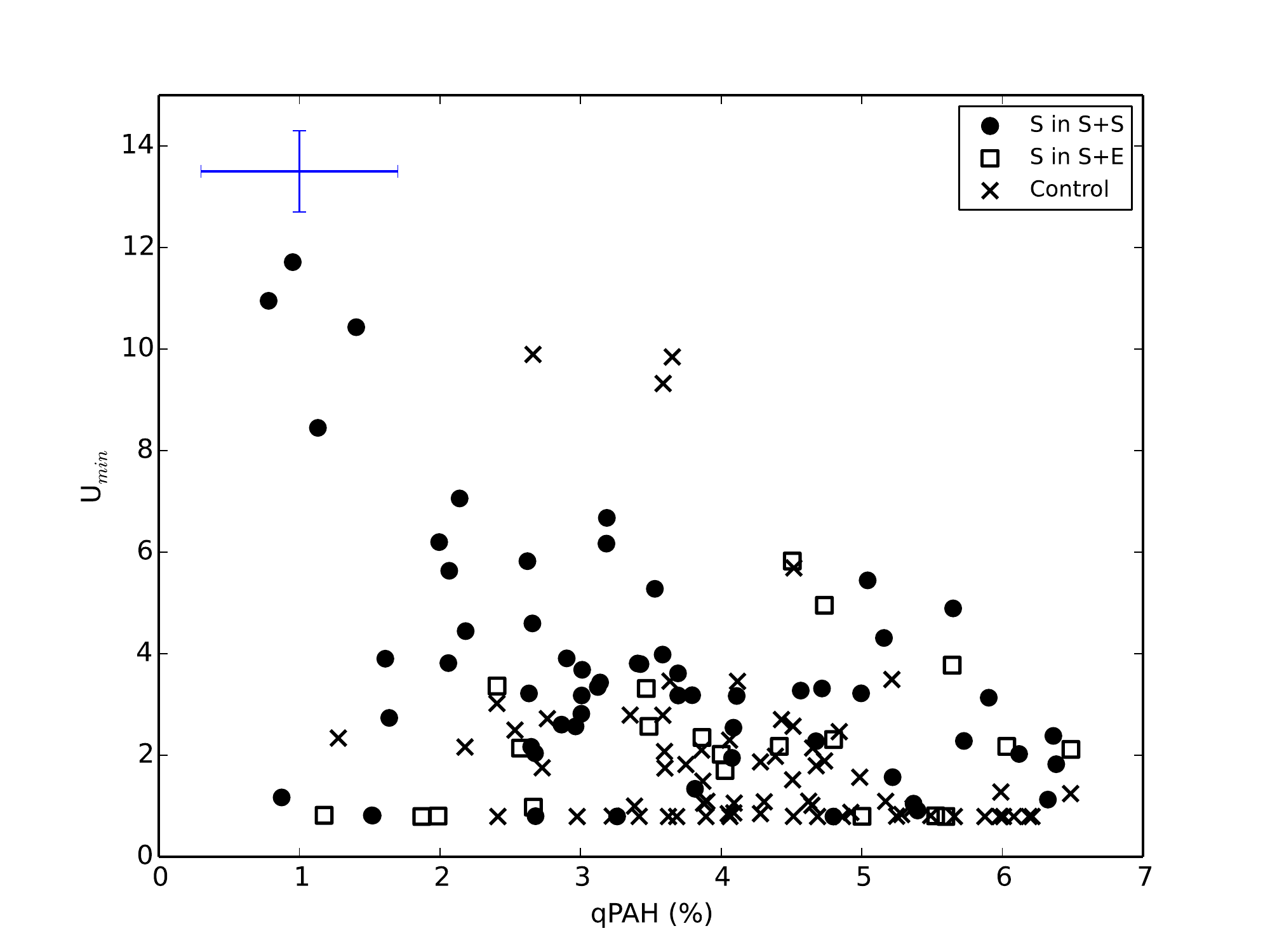}
\caption{U$_{min}$ vs. qPAH for the galaxy samples including spirals in S$+$S (circles), S$+$E (squares) and the control galaxies (cross). The blue error bars indicate the average error for each parameter.\label{fig:uminqPAH}
}
\end{figure}

\begin{figure}[!htb]
\includegraphics[width=\textwidth,angle=0]{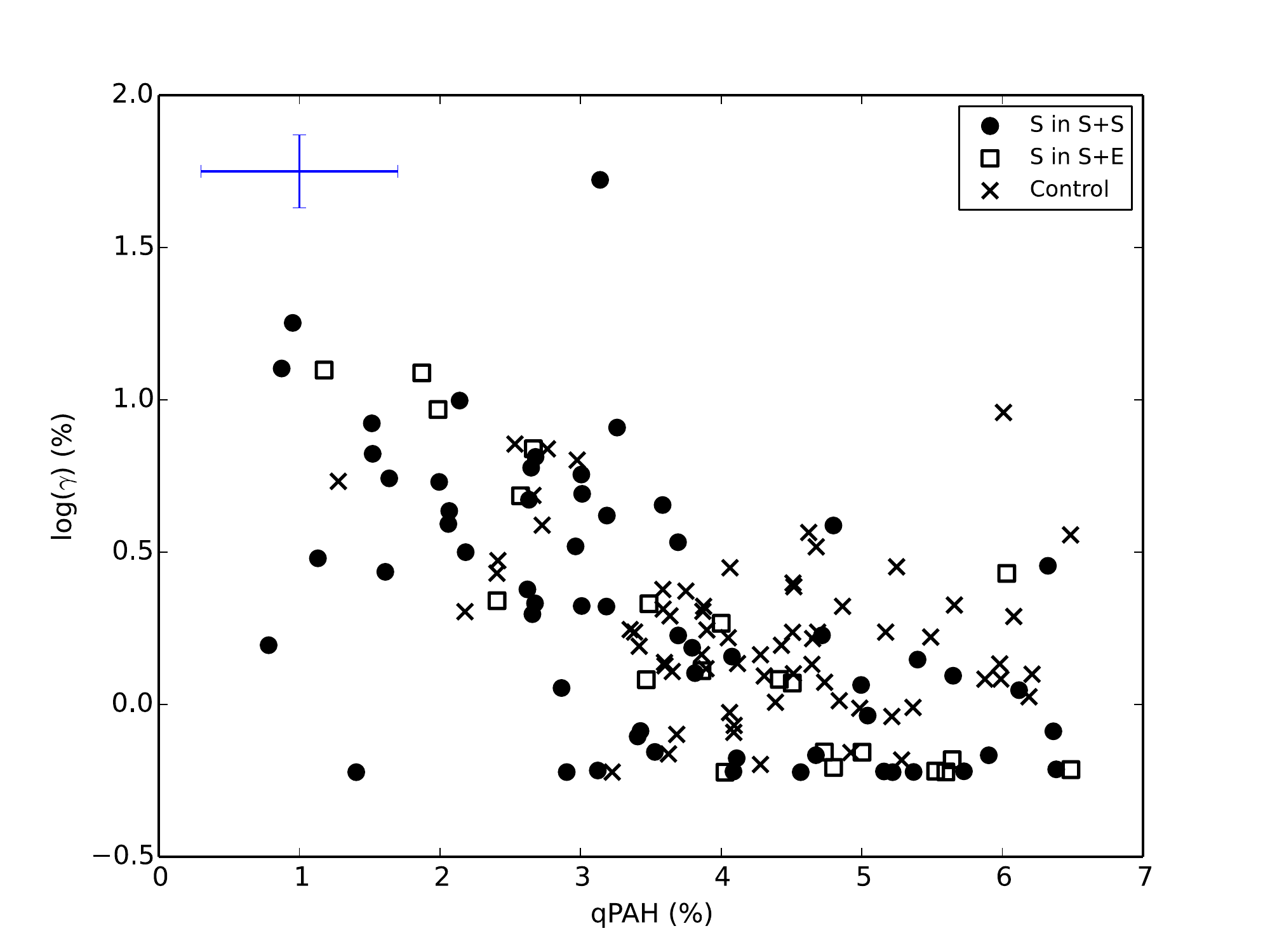}
\caption{log($\gamma$ (\%))  vs. qPAH for the galaxy samples including spirals in S$+$S (circles), S$+$E (squares) and the control galaxies (cross). The blue error bars indicate the average error for each parameter.\label{fig:gammaqPAH}
}
\end{figure}

\begin{figure}[!htb]
\includegraphics[width=\textwidth,angle=0]{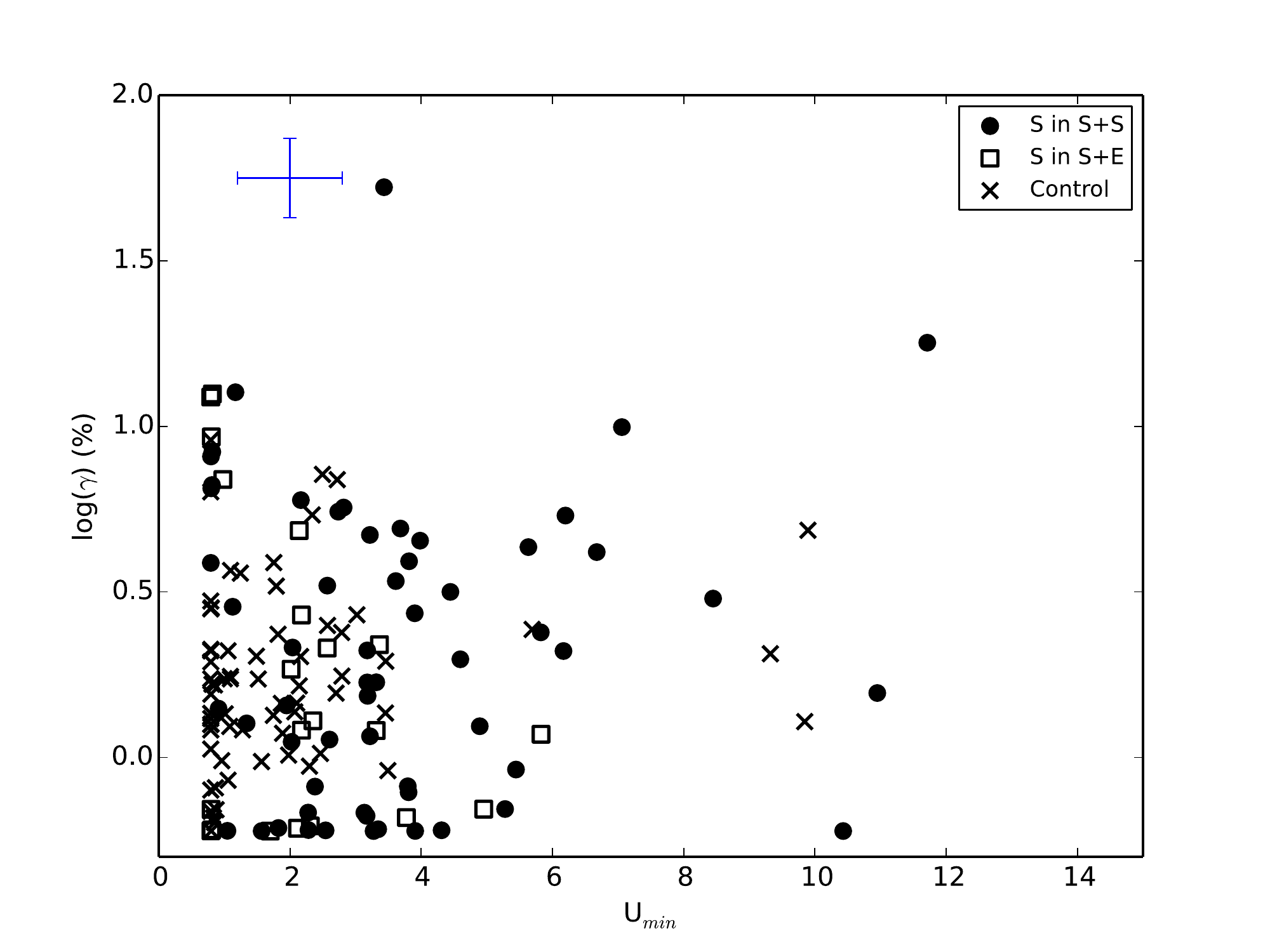}
\caption{log($\gamma$ (\%))  vs. U$_{min}$ for the galaxy samples including spirals in S$+$S (circles), S$+$E (squares) and the control galaxies (cross). The blue error bars indicate the average error for each parameter.\label{fig:gammaUmin}
}
\end{figure}

In order to investigate the underlying conditions for the DL07 dust parameters, we examine the correlations of U$_{min}$, qPAH and $\gamma$ with respect to each other. Fig.\ref{fig:uminqPAH} shows that there is no significant correlation of the qPAH and U$_{min}$ of any of the samples in our study. It can be seen from Fig.\ref{fig:uminqPAH} that spirals in S$+$S pairs have the highest U$_{min}$ with the most extreme cases possessing a very low qPAH. Our next comparison of the three DL07 parameters in Fig.\ref{fig:gammaUmin} does not display any significant correlation of $\gamma$ and U$_{min}$.
Fig.\ref{fig:gammaqPAH} illustrates the $\gamma$ vs. qPAH moderate and strong anti-correlations (Spearman $\rho$= -0.59, -0.76) for the spirals in S$+$S and S$+$E pairs, respectively. The anti-correlation for the control galaxies is indicated as weak. According to Fig.\ref{fig:gammaqPAH} paired spirals have a lower fraction of PAH (as measured by qPAH) when there is a larger fraction of dust heated above U$_{min}$ (as measured by $\gamma$). This is similar to the f12/f25 vs f60/f100 anti-correlation for iras galaxies. Low qPAH control galaxies follow this trend as well but the lack of sample correlation is due to the relatively higher qPAH expressed in the control group. 


\begin{figure}[!htb]
\includegraphics[width=\textwidth,angle=0]{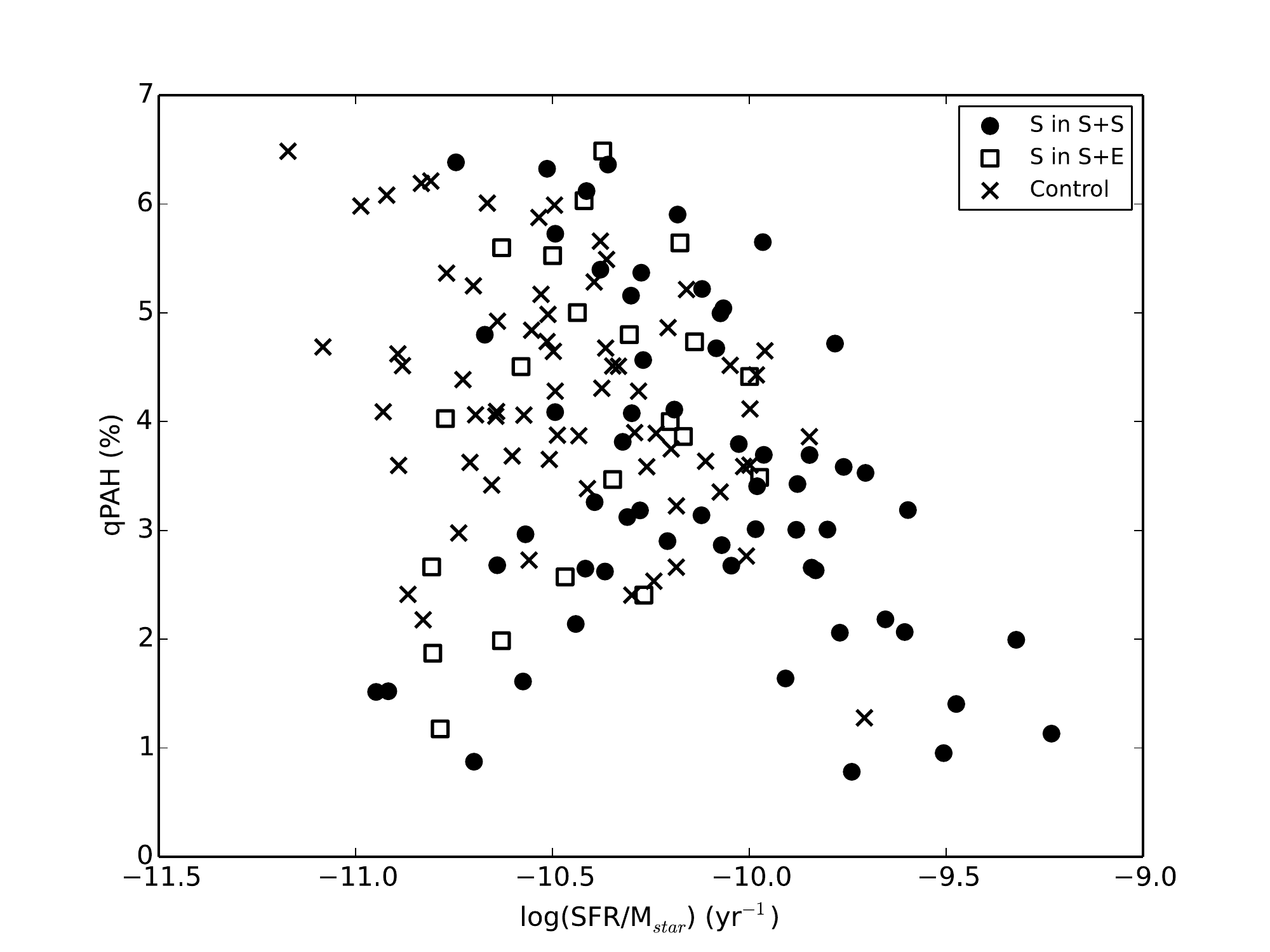}
\caption{The qPAH as a function of sSFR for spirals in S$+$S (circles), S$+$E (squares) and the control galaxies (crosses).\label{fig:ssfr_qpah}
}
\end{figure}

Enhancements in \citet{cao:2016} are best shown in the sSFR which may have an influence on the DL07 parameters. There is no significant dependance by Spearman coefficient of qPAH on the sSFR in Fig.\ref{fig:ssfr_qpah}. It should be noted however that there is a population of S$+$S spirals with high sSFR which have the lowest qPAH values. These stand as unique among the three samples. 

\begin{figure}[!htb]
\includegraphics[width=\textwidth,angle=0]{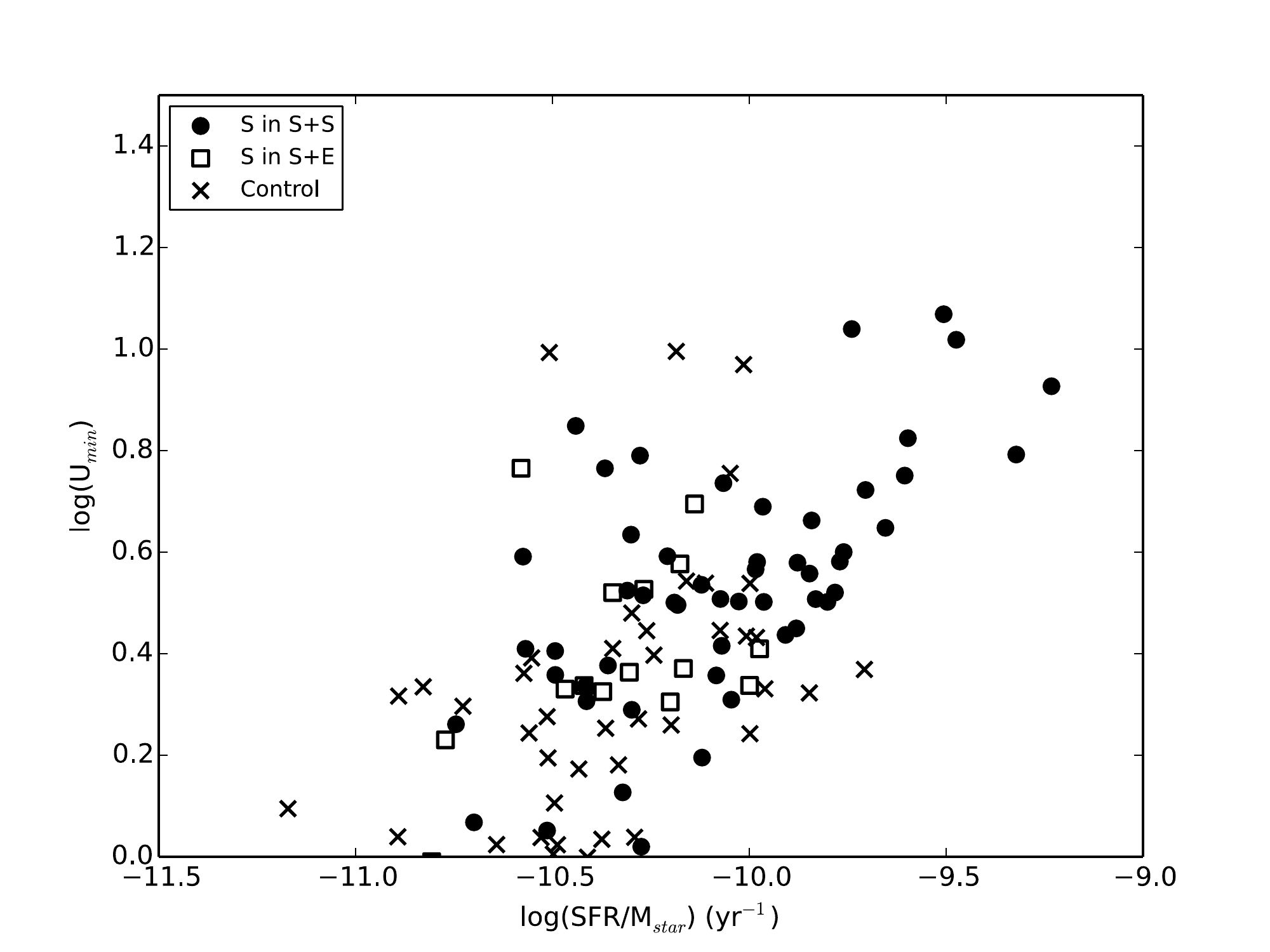}
\caption{The log(U$_{min}$) as a function of sSFR for spirals in S$+$S (circles), S$+$E (squares) and the control galaxies (crosses).\label{fig:ssfr_umin}
}
\end{figure}

All three samples, spirals in S$+$S, S$+$E, and control, show a strong correlation of U$_{min}$ with the sSFR (Fig.\ref{fig:ssfr_umin} (Spearman $\rho$=0.68, 0.67, 0.59 respectively). The upper range of sSFR and U$_{min}$ is dominated the by S$+$S spirals which have the low qPAH values seen in Fig.\ref{fig:ssfr_qpah}.

\section{Dust Parameter Enhancements}
\begin{figure}[!htb]
\includegraphics[width=\textwidth,angle=0]{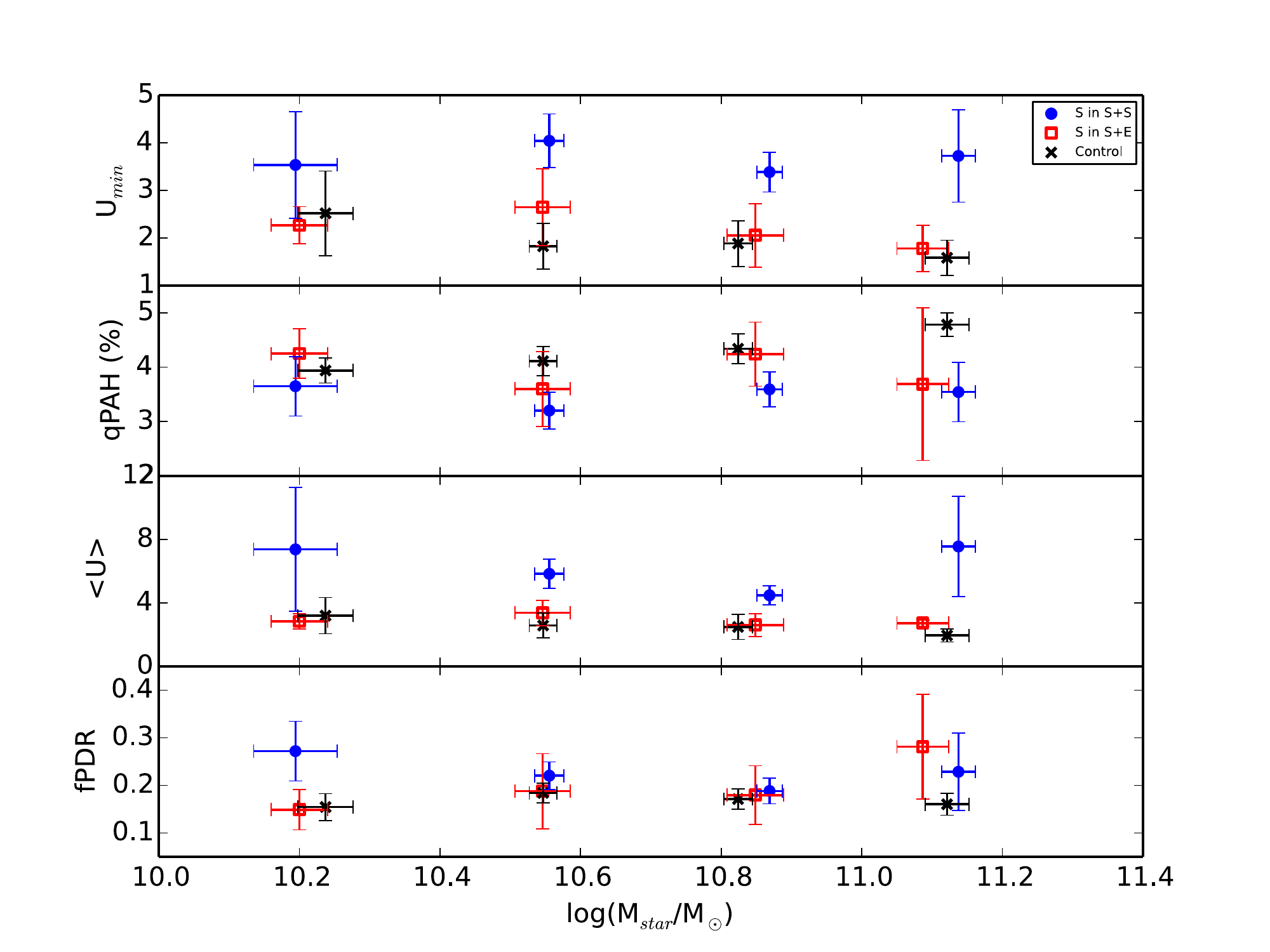}
\caption{The U$_{min}$, qPAH,$\langle$U$\rangle$, and fPDR for the galaxy samples including spirals in S$+$S (blue circles), S$+$E (red squares) and the control galaxies (black crosses). Galaxies are into divided into four mass bins and error bars represent the standard error\label{fig:massplot}.
}
\end{figure}

In order to determine the enhancement of the parameters which have different distributions from control in Fig.\ref{fig:histograms}, we create stellar mass bins as used in \citet{cao:2016}. The stellar mass bins are selected as log(M$_{star}$/M$_{\sun}$)$<$ 10.4, 10.4$\leq$log(M$_{star}$/M$_{\sun}$)$<$10.7,10.7 $\leq$log(M$_{star}$/M$_{\sun}$)$<$11.0, and log(M$_{star}$/M$_{\sun}$)$\geq$11.0. Errors based on binning are reported as standard errors (standard deviation of the mean) technique.  Fig.\ref{fig:massplot} displays the U$_{min}$, qPAH, $\langle$U$\rangle$, and fPDR for the spirals in S$+$S, S$+$E, and control samples. All 4 parameters for the S$+$E spirals are within the errors of the control sample. However, in contrast to the S$+$E pairs, the U$_{min}$ in S$+$S spirals significantly exceeds that of the control sample and the S$+$E in the three upper mass bins. The parameter U$_{min}$ is correlated to the overall dust temperature, T$_{dust}$, based on a power law relationship presented in the literature with varying but similar conversion factors \citep{Aniano:2012, Hunt:2015}.

The qPAH and $\langle$U$\rangle$ of S$+$S spirals is also seen to be significantly different from the parameters of the control in the same three upper mass bins as the excess in U$_{min}$. The qPAH is lower than both other samples but only significantly against the control. $\langle$U$\rangle$ exceeds both that of the spirals in S$+$E and control. The values of fPDR only differ in the lowest mass bin for S$+$S spirals vs. both other samples. There are no overall correlations of any of these parameters with stellar mass. The $\epsilon$ enhancements are defined as the difference of values averaged over all mass bins and their corresponding averages for the control sample (Table 3). The enhancements of U$_{min}$, $\langle$U$\rangle$ and reduction of qPAH in the S$+$S pairs is in sharp contrast to the similarity of the S$+$E to the control sample.

\begin{deluxetable}{lcccc}
\label{tbl:enhancements}
%\tabletypesize{\normalsize}
%\tabletypesize{\tiny}
\tabletypesize{\scriptsize}
%\tiny(5pt);\scriptsize(7pt);\footnotesize(8pt);\small(9pt);\normalsize(10pt)
\setlength{\tabcolsep}{0.05in} %Tighten up the columns. See AASTeX FAQ
%\rotate
\tablenum{3}
\tablewidth{0pt}
%\tablewidth{0.95\linewidth}
\tablecaption{Dust Parameter Enhancements}
\tablehead{
Sample&$\epsilon$$_{Umin}$&$\epsilon$$_{qPAH}$&$\epsilon$$_{\langle U \rangle}$&$\epsilon$$_{fPDR}$\\
}
\startdata
S$+$S&1.72$\pm$	0.55&	-0.79$\pm$	0.33&	3.19$\pm$	1.13&	0.04$\pm$	0.03\\
S$+$E &0.29$\pm$	0.54&	-0.25$\pm$	0.46&	0.31$\pm$	0.65&	0.02$\pm$	0.04\\

\enddata
%\tablecomments{Enhancement of the dust parameters over control sample.}
\end{deluxetable}

\section{Summary and Discussion}

We present a CIGALE analysis of the dust parameters defined in the DL07 dust models for a sample of close major-merger galaxy pairs. The sample morphology is both spiral-spiral (S$+$S) and spiral-elliptical (S$+$E) allowing for a probe into the interaction physics and its influence on dust. The contributions (qPAH) of PAH grains to the dust mass, fraction ($\gamma$) of dust from photodissociation regions (PDR) to the total dust and the minimum intensity (U$_{min}$) from the interstellar radiation field (ISRF) are analyzed along with their secondary related parameters such as mean radiation intensity, $\langle$U$\rangle$, and dust luminosity (total and in PDRs). 

\citet{Elbaz:2011} demonstrate the contribution of the diffuse ISM templates as compared to star forming templates within the GOODS-Herschel survey. The corresponding ISM $\langle$U$\rangle$, T$_{eff}$ and fPAH (the PAH-to-total mass fraction; compare to qPAH) are given as 1.8, 19 K and 8.74\% respectively. Star-forming regions are measured as having $\langle$U$\rangle$, T$_{eff}$ and fPAH as 757, 53 K and 1.38\%. Galaxies exhibit a mixture of the two templates and our H-KPAIR parameters  give us the ability to access to relative importance of ISM and star-forming regions to our paired spirals. The \citet{Elbaz:2011} analysis would imply the H-KPAIR spirals in S$+$S pairs have the largest fraction of star-forming regions based on temperature, $\langle$U$\rangle$ and qPAH as seen in Fig.\ref{fig:massplot} among our samples. 

Any display of smaller qPAH such as the reduction seen in the S$+$S pairs is likely a PAH deficiency due to enhanced interstellar radiation fields (ISRF) from star formation in the pair environment \citep{Contursi:2000,cook:2014, madden:2006, Engelbracht:2005}. The main-sequence of galaxies \citep{Elbaz:2011} have a log(sSFR)$\sim$-9.6 with star-bursting galaxies defined as log(sSFR)$\sim$-9.3. \citet{Nordon:2012} demonstrate that the difference in sSFR from the expected main-sequence \citep{Elbaz:2011} of galaxies ($\Delta$sSFR$_{MS}$) is a determinant in the relative strength of the luminosity at 8$\mu$m which diminishes at higher $\Delta$sSFR$_{MS}$.  The 8$\mu$m range contains the PAH feature measured in the corresponding Spitzer IRAC band and a reduced qPAH would likely be a measurable result of the same physical environment. The population of S$+$S spirals has an enhanced $\Delta$sSFR$_{MS}$ according to their sSFR enhancement\citep{cao:2016}. Since qPAH is diminished for a population of these spirals with log(sSFR)$>$ -10 (Fig.\ref{fig:ssfr_qpah}), it is likely a reflection of the same shift in SED shape as seen in \citet{Nordon:2012} which are interpreted as the result more compact star-forming regions\citep{Elbaz:2011}. The low qPAH galaxy examples have sSFR which falls in range from main-sequence to just below that of the star-bursting definition. The ``compactness" defined in \citet{Elbaz:2011} refers an overall measure of extended vs. non-extended location of star-bursting and not to the details of the size of individual regions or PDRs. So despite an indication of more compact star formation, the $\gamma$ parameter shows a strong anti-correlation with qPAH (Fig.\ref{fig:gammaqPAH}) in these same galaxies, an indication of a larger fraction of dust heated above the diffuse interstellar medium temperature.

The similarity of the spirals in the S$+$E pairs to control spirals indicate a difference in the interaction physics or history of the S$+$E pairs when compared to S$+$S pairs. Possible conditions which could cause the lack of enhancement are the presence of multiple star forming mechanisms in S$+$S pairs such as torque induced gas flow as well as cloud-cloud interactions and turbulence \citep{Bournaud:2011,Renaud:2014}. The S$+$E pairs either only have the torque induced mechanism at their disposal because of the lack of two gas rich environments or the presence of a hot intergalactic medium (IGM) associated with the elliptical suppresses the star formation or both suppression conditions are present.

\acknowledgments 
This publication makes use of data products from the Wide-field Infrared Survey Explorer, which is a joint project of the University of California, Los Angeles, and the Jet Propulsion Laboratory/California Institute of Technology, funded by the National Aeronautics and Space Administration.
The Herschel spacecraft was designed, built, tested, and launched under a contract to ESA man- aged by the Herschel/Planck Project team by an industrial consortium under the overall responsibility of the prime con- tractor Thales Alenia Space (Cannes), and including Astrium (Friedrichshafen) responsible for the payload module and for system testing at spacecraft level, Thales Alenia Space (Turin) responsible for the service module, and Astrium (Toulouse) responsible for the telescope, with in excess of a hundred sub- contractors. C.C. is supported by NSFC-11503013, NSFC-11420101002, and NSFC-10978014. We would also like to thank the anonymous reviewer for improvements to this manuscript.

\bibliographystyle{apj}
\bibliography{my_bib}

\begin{thebibliography}{}
\expandafter\ifx\csname natexlab\endcsname\relax\def\natexlab#1{#1}\fi

\bibitem[{{Adelman-McCarthy} {et~al.}(2007)}]{Adelman-McCarthy:2007}
{Adelman-McCarthy}, J.~K., {et~al.} 2007, \apjs, 172, 634

\bibitem[{{Alonso-Herrero} {et~al.}(2014){Alonso-Herrero}, {Ramos Almeida},
  {Esquej}, {Roche}, {Hern{\'a}n-Caballero}, {H{\"o}nig},
  {Gonz{\'a}lez-Mart{\'{\i}}n}, {Aretxaga}, {Mason}, {Packham}, {Levenson},
  {Rodr{\'{\i}}guez Espinosa}, {Siebenmorgen}, {Pereira-Santaella},
  {D{\'{\i}}az-Santos}, {Colina}, {Alvarez}, \& {Telesco}}]{Alonso:AGNPAH}
{Alonso-Herrero}, A., {Ramos Almeida}, C., {Esquej}, P., {et~al.} 2014, \mnras,
  443, 2766

\bibitem[{{Aniano} {et~al.}(2012){Aniano}, {Draine}, {Calzetti}, {Dale},
  {Engelbracht}, {Gordon}, {Hunt}, {Kennicutt}, {Krause}, {Leroy}, {Rix},
  {Roussel}, {Sandstrom}, {Sauvage}, {Walter}, {Armus}, {Bolatto}, {Crocker},
  {Donovan Meyer}, {Galametz}, {Helou}, {Hinz}, {Johnson}, {Koda}, {Montiel},
  {Murphy}, {Skibba}, {Smith}, \& {Wolfire}}]{Aniano:2012}
{Aniano}, G., {Draine}, B.~T., {Calzetti}, D., {et~al.} 2012, \apj, 756, 138

\bibitem[{{Barton} {et~al.}(2007){Barton}, {Arnold}, {Zentner}, {Bullock}, \&
  {Wechsler}}]{Barton:2007}
{Barton}, E.~J., {Arnold}, J.~A., {Zentner}, A.~R., {Bullock}, J.~S., \&
  {Wechsler}, R.~H. 2007, \apj, 671, 1538

\bibitem[{{Barton} {et~al.}(2000){Barton}, {Geller}, \& {Kenyon}}]{Barton:2000}
{Barton}, E.~J., {Geller}, M.~J., \& {Kenyon}, S.~J. 2000, \apj, 530, 660

\bibitem[{{Bournaud}(2011)}]{Bournaud:2011}
{Bournaud}, F. 2011, in EAS Publications Series, Vol.~51, EAS Publications
  Series, ed. C.~{Charbonnel} \& T.~{Montmerle}, 107--131

\bibitem[{{Brinchmann} {et~al.}(1998){Brinchmann}, {Abraham}, {Schade},
  {Tresse}, {Ellis}, {Lilly}, {Le Fevre}, {Glazebrook}, {Hammer}, {Colless},
  {Crampton}, \& {Broadhurst}}]{Brinchmann:1998}
{Brinchmann}, J., {Abraham}, R., {Schade}, D., {et~al.} 1998, \apj, 499, 112

\bibitem[{{Bundy} {et~al.}(2004){Bundy}, {Fukugita}, {Ellis}, {Kodama}, \&
  {Conselice}}]{Bundy:2004}
{Bundy}, K., {Fukugita}, M., {Ellis}, R.~S., {Kodama}, T., \& {Conselice},
  C.~J. 2004, \apjl, 601, L123

\bibitem[{Bundy {et~al.}(2009)Bundy, Fukugita, Ellis, Targett, Belli, \&
  Kodama}]{Bundy:2009}
Bundy, K., Fukugita, M., Ellis, R.~S., {et~al.} 2009, The Astrophysical
  Journal, 697, 1369

\bibitem[{{Burkert}(2006)}]{Burkert:2006}
{Burkert}, A. 2006, Comptes Rendus Physique, 7, 433

\bibitem[{{Cao} {et~al.}(2016){Cao}, {Xu}, {Domingue}, {Buat}, {Cheng}, {Gao},
  {Huang}, {Jarrett}, {Lisenfeld}, {Lu}, {Mazzarella}, {Sun}, {Wu}, {Yun},
  {Ronca}, \& {Jacques}}]{cao:2016}
{Cao}, C., {Xu}, C.~K., {Domingue}, D., {et~al.} 2016, \apjs, 222, 16

\bibitem[{{Ciesla} {et~al.}(2014){Ciesla}, {Boquien}, {Boselli}, {Buat},
  {Cortese}, {Bendo}, {Heinis}, {Galametz}, {Eales}, {Smith}, {Baes},
  {Bianchi}, {De Looze}, {di Serego Alighieri}, {Galliano}, {Hughes}, {Madden},
  {Pierini}, {R{\'e}my-Ruyer}, {Spinoglio}, {Vaccari}, {Viaene}, \&
  {Vlahakis}}]{Ciesla:2014}
{Ciesla}, L., {Boquien}, M., {Boselli}, A., {et~al.} 2014, \aap, 565, A128

\bibitem[{{Cluver} {et~al.}(2014){Cluver}, {Jarrett}, {Hopkins}, {Driver},
  {Liske}, {Gunawardhana}, {Taylor}, {Robotham}, {Alpaslan}, {Baldry}, {Brown},
  {Peacock}, {Popescu}, {Tuffs}, {Bauer}, {Bland-Hawthorn}, {Colless},
  {Holwerda}, {Lara-L{\'o}pez}, {Leschinski}, {L{\'o}pez-S{\'a}nchez},
  {Norberg}, {Owers}, {Wang}, \& {Wilkins}}]{GAMA:WISE}
{Cluver}, M.~E., {Jarrett}, T.~H., {Hopkins}, A.~M., {et~al.} 2014, \apj, 782,
  90

\bibitem[{{Cole} {et~al.}(2001)}]{Cole:2001}
{Cole}, S., {et~al.} 2001, \mnras, 326, 255

\bibitem[{{Conselice} {et~al.}(2003){Conselice}, {Bershady}, {Dickinson}, \&
  {Papovich}}]{Conselice:2003}
{Conselice}, C.~J., {Bershady}, M.~A., {Dickinson}, M., \& {Papovich}, C. 2003,
  \aj, 126, 1183

\bibitem[{{Contursi} {et~al.}(2000){Contursi}, {Lequeux}, {Cesarsky},
  {Boulanger}, {Rubio}, {Hanus}, {Sauvage}, {Tran}, {Bosma}, {Madden}, \&
  {Vigroux}}]{Contursi:2000}
{Contursi}, A., {Lequeux}, J., {Cesarsky}, D., {et~al.} 2000, \aap, 362, 310

\bibitem[{{Cook} {et~al.}(2014){Cook}, {Dale}, {Johnson}, {Van Zee}, {Lee},
  {Kennicutt}, {Calzetti}, {Staudaher}, \& {Engelbracht}}]{cook:2014}
{Cook}, D.~O., {Dale}, D.~A., {Johnson}, B.~D., {et~al.} 2014, \mnras, 445, 899

\bibitem[{{Cutri} \& {et al.}(2013)}]{AllWISE}
{Cutri}, R.~M., \& {et al.} 2013, VizieR Online Data Catalog, 2328, 0

\bibitem[{{Dale} {et~al.}(2012){Dale}, {Aniano}, {Engelbracht}, {Hinz},
  {Krause}, {Montiel}, {Roussel}, {Appleton}, {Armus}, {Beir{\~a}o}, {Bolatto},
  {Brandl}, {Calzetti}, {Crocker}, {Croxall}, {Draine}, {Galametz}, {Gordon},
  {Groves}, {Hao}, {Helou}, {Hunt}, {Johnson}, {Kennicutt}, {Koda}, {Leroy},
  {Li}, {Meidt}, {Miller}, {Murphy}, {Rahman}, {Rix}, {Sandstrom}, {Sauvage},
  {Schinnerer}, {Skibba}, {Smith}, {Tabatabaei}, {Walter}, {Wilson}, {Wolfire},
  \& {Zibetti}}]{Dale2012}
{Dale}, D.~A., {Aniano}, G., {Engelbracht}, C.~W., {et~al.} 2012, \apj, 745, 95

\bibitem[{Darg {et~al.}(2010)Darg, Kaviraj, Lintott, Schawinski, Sarzi,
  Bamford, Silk, Proctor, Andreescu, Murray, Nichol, Raddick, Slosar, Szalay,
  Thomas, \& Vandenberg}]{Darg2010}
Darg, D.~W., Kaviraj, S., Lintott, C.~J., {et~al.} 2010, Monthly Notices of the
  Royal Astronomical Society, 401, 1043

\bibitem[{{Dasyra} {et~al.}(2006){Dasyra}, {Tacconi}, {Davies}, {Genzel},
  {Lutz}, {Naab}, {Burkert}, {Veilleux}, \& {Sanders}}]{Dasyra:2006}
{Dasyra}, K.~M., {Tacconi}, L.~J., {Davies}, R.~I., {et~al.} 2006, \apj, 638,
  745

\bibitem[{{Di Matteo} {et~al.}(2008){Di Matteo}, {Bournaud}, {Martig},
  {Combes}, {Melchior}, \& {Semelin}}]{Di-Matteo:2008}
{Di Matteo}, P., {Bournaud}, F., {Martig}, M., {et~al.} 2008, \aap, 492, 31

\bibitem[{{Domingue} {et~al.}(2009){Domingue}, {Xu}, {Jarrett}, \&
  {Cheng}}]{Domingue:2009mq}
{Domingue}, D.~L., {Xu}, C.~K., {Jarrett}, T.~H., \& {Cheng}, Y. 2009, \apj,
  695, 1559

\bibitem[{{Draine} \& {Lee}(1984)}]{Draine:1984}
{Draine}, B.~T., \& {Lee}, H.~M. 1984, \apj, 285, 89

\bibitem[{{Draine} \& {Li}(2007)}]{Draine:2007ta}
{Draine}, B.~T., \& {Li}, A. 2007, \apj, 657, 810

\bibitem[{{Draine} {et~al.}(2007){Draine}, {Dale}, {Bendo}, {Gordon}, {Smith},
  {Armus}, {Engelbracht}, {Helou}, {Kennicutt}, {Li}, {Roussel}, {Walter},
  {Calzetti}, {Moustakas}, {Murphy}, {Rieke}, {Bot}, {Hollenbach}, {Sheth}, \&
  {Teplitz}}]{DraineDale:2007}
{Draine}, B.~T., {Dale}, D.~A., {Bendo}, G., {et~al.} 2007, \apj, 663, 866

\bibitem[{{Eisenstein} {et~al.}(2011){Eisenstein}, {Weinberg}, {Agol},
  {Aihara}, {Allende Prieto}, {Anderson}, {Arns}, {Aubourg}, {Bailey},
  {Balbinot}, \& et~al.}]{SDSSIII}
{Eisenstein}, D.~J., {Weinberg}, D.~H., {Agol}, E., {et~al.} 2011, \aj, 142, 72

\bibitem[{{Elbaz} {et~al.}(2011){Elbaz}, {Dickinson, M.}, {Hwang, H. S.},
  {D{\'\i}az-Santos, T.}, {Magdis, G.}, {Magnelli, B.}, {Le Borgne, D.},
  {Galliano, F.}, {Pannella, M.}, {Chanial, P.}, {Armus, L.}, {Charmandaris,
  V.}, {Daddi, E.}, {Aussel, H.}, {Popesso, P.}, {Kartaltepe, J.}, {Altieri,
  B.}, {Valtchanov, I.}, {Coia, D.}, {Dannerbauer, H.}, {Dasyra, K.}, {Leiton,
  R.}, {Mazzarella, J.}, {Alexander, D. M.}, {Buat, V.}, {Burgarella, D.},
  {Chary, R.-R.}, {Gilli, R.}, {Ivison, R. J.}, {Juneau, S.}, {Le Floc'h, E.},
  {Lutz, D.}, {Morrison, G. E.}, {Mullaney, J. R.}, {Murphy, E.}, {Pope, A.},
  {Scott, D.}, {Brodwin, M.}, {Calzetti, D.}, {Cesarsky, C.}, {Charlot, S.},
  {Dole, H.}, {Eisenhardt, P.}, {Ferguson, H. C.}, {F{\"o}rster Schreiber, N.},
  {Frayer, D.}, {Giavalisco, M.}, {Huynh, M.}, {Koekemoer, A. M.}, {Papovich,
  C.}, {Reddy, N.}, {Surace, C.}, {Teplitz, H.}, {Yun, M. S.}, \& {Wilson,
  G.}}]{Elbaz:2011}
{Elbaz}, D., {Dickinson, M.}, {Hwang, H. S.}, {et~al.} 2011, \aap, 533, A119

\bibitem[{{Elmegreen} {et~al.}(1995){Elmegreen}, {Kaufman}, {Brinks},
  {Elmegreen}, \& {Sundin}}]{Elmegreen:1995}
{Elmegreen}, D.~M., {Kaufman}, M., {Brinks}, E., {Elmegreen}, B.~G., \&
  {Sundin}, M. 1995, \apj, 453, 100

\bibitem[{{Engelbracht} {et~al.}(2005){Engelbracht}, {Gordon}, {Rieke},
  {Werner}, {Dale}, \& {Latter}}]{Engelbracht:2005}
{Engelbracht}, C.~W., {Gordon}, K.~D., {Rieke}, G.~H., {et~al.} 2005, \apjl,
  628, L29

\bibitem[{{Erwin}(2015)}]{IMFIT}
{Erwin}, P. 2015, \apj, 799, 226

\bibitem[{{Griffin} {et~al.}(2010){Griffin}, {Abergel}, {Abreu}, {Ade},
  {Andr{\'e}}, {Augueres}, {Babbedge}, {Bae}, {Baillie}, {Baluteau}, {Barlow},
  {Bendo}, {Benielli}, {Bock}, {Bonhomme}, {Brisbin}, {Brockley-Blatt},
  {Caldwell}, {Cara}, {Castro-Rodriguez}, {Cerulli}, {Chanial}, {Chen},
  {Clark}, {Clements}, {Clerc}, {Coker}, {Communal}, {Conversi}, {Cox},
  {Crumb}, {Cunningham}, {Daly}, {Davis}, {de Antoni}, {Delderfield}, {Devin},
  {di Giorgio}, {Didschuns}, {Dohlen}, {Donati}, {Dowell}, {Dowell}, {Duband},
  {Dumaye}, {Emery}, {Ferlet}, {Ferrand}, {Fontignie}, {Fox}, {Franceschini},
  {Frerking}, {Fulton}, {Garcia}, {Gastaud}, {Gear}, {Glenn}, {Goizel},
  {Griffin}, {Grundy}, {Guest}, {Guillemet}, {Hargrave}, {Harwit}, {Hastings},
  {Hatziminaoglou}, {Herman}, {Hinde}, {Hristov}, {Huang}, {Imhof}, {Isaak},
  {Israelsson}, {Ivison}, {Jennings}, {Kiernan}, {King}, {Lange}, {Latter},
  {Laurent}, {Laurent}, {Leeks}, {Lellouch}, {Levenson}, {Li}, {Li},
  {Lilienthal}, {Lim}, {Liu}, {Lu}, {Madden}, {Mainetti}, {Marliani}, {McKay},
  {Mercier}, {Molinari}, {Morris}, {Moseley}, {Mulder}, {Mur}, {Naylor},
  {Nguyen}, {O'Halloran}, {Oliver}, {Olofsson}, {Olofsson}, {Orfei}, {Page},
  {Pain}, {Panuzzo}, {Papageorgiou}, {Parks}, {Parr-Burman}, {Pearce},
  {Pearson}, {P{\'e}rez-Fournon}, {Pinsard}, {Pisano}, {Podosek}, {Pohlen},
  {Polehampton}, {Pouliquen}, {Rigopoulou}, {Rizzo}, {Roseboom}, {Roussel},
  {Rowan-Robinson}, {Rownd}, {Saraceno}, {Sauvage}, {Savage}, {Savini},
  {Sawyer}, {Scharmberg}, {Schmitt}, {Schneider}, {Schulz}, {Schwartz},
  {Shafer}, {Shupe}, {Sibthorpe}, {Sidher}, {Smith}, {Smith}, {Smith},
  {Spencer}, {Stobie}, {Sudiwala}, {Sukhatme}, {Surace}, {Stevens}, {Swinyard},
  {Trichas}, {Tourette}, {Triou}, {Tseng}, {Tucker}, {Turner}, {Vaccari},
  {Valtchanov}, {Vigroux}, {Virique}, {Voellmer}, {Walker}, {Ward}, {Waskett},
  {Weilert}, {Wesson}, {White}, {Whitehouse}, {Wilson}, {Winter}, {Woodcraft},
  {Wright}, {Xu}, {Zavagno}, {Zemcov}, {Zhang}, \& {Zonca}}]{SPIRE:2010}
{Griffin}, M.~J., {Abergel}, A., {Abreu}, A., {et~al.} 2010, \aap, 518, L3

\bibitem[{{Helou} {et~al.}(2000){Helou}, {Lu}, {Werner}, {Malhotra}, \&
  {Silbermann}}]{Helou:2000}
{Helou}, G., {Lu}, N.~Y., {Werner}, M.~W., {Malhotra}, S., \& {Silbermann}, N.
  2000, \apjl, 532, L21

\bibitem[{Hopkins {et~al.}(2006)Hopkins, Hernquist, Cox, Matteo, Robertson, \&
  Springel}]{Hopkins:2006}
Hopkins, P.~F., Hernquist, L., Cox, T.~J., {et~al.} 2006, The Astrophysical
  Journal Supplement Series, 163, 1

\bibitem[{{Hunt} {et~al.}(2015){Hunt}, {Draine}, {Bianchi}, {Gordon}, {Aniano},
  {Calzetti}, {Dale}, {Helou}, {Hinz}, {Kennicutt}, {Roussel}, {Wilson},
  {Bolatto}, {Boquien}, {Croxall}, {Galametz}, {Gil de Paz}, {Koda},
  {Mu{\~n}oz-Mateos}, {Sandstrom}, {Sauvage}, {Vigroux}, \&
  {Zibetti}}]{Hunt:2015}
{Hunt}, L.~K., {Draine}, B.~T., {Bianchi}, S., {et~al.} 2015, \aap, 576, A33

\bibitem[{{Hwang} {et~al.}(2011){Hwang}, {Elbaz}, {Dickinson}, {Charmandaris},
  {Daddi}, {Le Borgne}, {Buat}, {Magdis}, {Altieri}, {Aussel}, {Coia},
  {Dannerbauer}, {Dasyra}, {Kartaltepe}, {Leiton}, {Magnelli}, {Popesso}, \&
  {Valtchanov}}]{Hwang:2011qw}
{Hwang}, H.~S., {Elbaz}, D., {Dickinson}, M., {et~al.} 2011, \aap, 535, A60

\bibitem[{{Jarrett} {et~al.}(2000){Jarrett}, {Chester}, {Cutri}, {Schneider},
  {Skrutskie}, \& {Huchra}}]{Jarrett:2000}
{Jarrett}, T.~H., {Chester}, T., {Cutri}, R., {et~al.} 2000, \aj, 119, 2498

\bibitem[{{Jarrett} {et~al.}(2011){Jarrett}, {Cohen}, {Masci}, {Wright},
  {Stern}, {Benford}, {Blain}, {Carey}, {Cutri}, {Eisenhardt}, {Lonsdale},
  {Mainzer}, {Marsh}, {Padgett}, {Petty}, {Ressler}, {Skrutskie}, {Stanford},
  {Surace}, {Tsai}, {Wheelock}, \& {Yan}}]{jarrett:2011}
{Jarrett}, T.~H., {Cohen}, M., {Masci}, F., {et~al.} 2011, \apj, 735, 112

\bibitem[{Jarrett {et~al.}(2012)Jarrett, Masci, Tsai, Petty, Cluver, Assef,
  Benford, Blain, Bridge, Donoso, Eisenhardt, Fowler, Koribalski, Lake, Neill,
  Seibert, Sheth, Stanford, \& Wright}]{jarrett:2012}
Jarrett, T.~H., Masci, F., Tsai, C.~W., {et~al.} 2012, /aj, 144, 68

\bibitem[{{Jarrett} {et~al.}(2013){Jarrett}, {Masci}, {Tsai}, {Petty},
  {Cluver}, {Assef}, {Benford}, {Blain}, {Bridge}, {Donoso}, {Eisenhardt},
  {Koribalski}, {Lake}, {Neill}, {Seibert}, {Sheth}, {Stanford}, \&
  {Wright}}]{jarrett:2013}
{Jarrett}, T.~H., {Masci}, F., {Tsai}, C.~W., {et~al.} 2013, \aj, 145, 6

\bibitem[{{Kauffmann} {et~al.}(1993){Kauffmann}, {White}, \&
  {Guiderdoni}}]{Kauffmann:1993}
{Kauffmann}, G., {White}, S.~D.~M., \& {Guiderdoni}, B. 1993, \mnras, 264, 201

\bibitem[{{Kennicutt} {et~al.}(1987){Kennicutt}, {Roettiger}, {Keel}, {van der
  Hulst}, \& {Hummel}}]{Kennicutt:1987}
{Kennicutt}, Jr., R.~C., {Roettiger}, K.~A., {Keel}, W.~C., {van der Hulst},
  J.~M., \& {Hummel}, E. 1987, \aj, 93, 1011

\bibitem[{{Kroupa}(2001)}]{Kroupa:IMF}
{Kroupa}, P. 2001, \mnras, 322, 231

\bibitem[{{Laher}(2012)}]{laher:2012}
{Laher}, R. 2012, {APT: Aperture Photometry Tool}, Astrophysics Source Code
  Library, ascl:1208.003

\bibitem[{{Lang}(2014)}]{unwise:images}
{Lang}, D. 2014, \aj, 147, 108

\bibitem[{{Lang} {et~al.}(2014){Lang}, {Hogg}, \& {Schlegel}}]{unwise:data}
{Lang}, D., {Hogg}, D.~W., \& {Schlegel}, D.~J. 2014, ArXiv e-prints,
  arXiv:1410.7397

\bibitem[{{Larson} \& {Tinsley}(1978)}]{Larson:1977}
{Larson}, R.~B., \& {Tinsley}, B.~M. 1978, \apj, 219, 46

\bibitem[{{Madden} {et~al.}(2006){Madden}, {Galliano}, {Jones}, \&
  {Sauvage}}]{madden:2006}
{Madden}, S.~C., {Galliano}, F., {Jones}, A.~P., \& {Sauvage}, M. 2006, \aap,
  446, 877

\bibitem[{{Maraston}(2005)}]{maraston:ssp}
{Maraston}, C. 2005, \mnras, 362, 799

\bibitem[{{Masci} \& {Fowler}(2009)}]{Masci:2009}
{Masci}, F.~J., \& {Fowler}, J.~W. 2009, in Astronomical Society of the Pacific
  Conference Series, Vol. 411, Astronomical Data Analysis Software and Systems
  XVIII, ed. D.~A. {Bohlender}, D.~{Durand}, \& P.~{Dowler}, 67

\bibitem[{{Mateos} {et~al.}(2012){Mateos}, {Alonso-Herrero}, {Carrera},
  {Blain}, {Watson}, {Barcons}, {Braito}, {Severgnini}, {Donley}, \&
  {Stern}}]{mateos:2012}
{Mateos}, S., {Alonso-Herrero}, A., {Carrera}, F.~J., {et~al.} 2012, \mnras,
  426, 3271

\bibitem[{{Mathis} {et~al.}(1977){Mathis}, {Rumpl}, \&
  {Nordsieck}}]{Mathis:1977}
{Mathis}, J.~S., {Rumpl}, W., \& {Nordsieck}, K.~H. 1977, \apj, 217, 425

\bibitem[{{Moon} \& {Yoon}(2015)}]{Moon:2015}
{Moon}, J.-S., \& {Yoon}, S.-J. 2015, Publication of Korean Astronomical
  Society, 30, 469

\bibitem[{{Noll} {et~al.}(2009){Noll}, {Burgarella}, {Giovannoli}, {Buat},
  {Marcillac}, \& {Mu{\~n}oz-Mateos}}]{Noll:2009}
{Noll}, S., {Burgarella}, D., {Giovannoli}, E., {et~al.} 2009, \aap, 507, 1793

\bibitem[{{Nordon} {et~al.}(2012){Nordon}, {Lutz}, {Genzel}, {Berta}, {Wuyts},
  {Magnelli}, {Altieri}, {Andreani}, {Aussel}, {Bongiovanni}, {Cepa},
  {Cimatti}, {Daddi}, {Fadda}, {F{\"o}rster Schreiber}, {Lagache}, {Maiolino},
  {P{\'e}rez Garc{\'{\i}}a}, {Poglitsch}, {Popesso}, {Pozzi}, {Rodighiero},
  {Rosario}, {Saintonge}, {Sanchez-Portal}, {Santini}, {Sturm}, {Tacconi},
  {Valtchanov}, \& {Yan}}]{Nordon:2012}
{Nordon}, R., {Lutz}, D., {Genzel}, R., {et~al.} 2012, \apj, 745, 182

\bibitem[{{O'Donnell} \& {Mathis}(1997)}]{Odonnell:1997}
{O'Donnell}, J.~E., \& {Mathis}, J.~S. 1997, \apj, 479, 806

\bibitem[{{Oliver} {et~al.}(2012){Oliver}, {Bock}, {Altieri}, {Amblard},
  {Arumugam}, {Aussel}, {Babbedge}, {Beelen}, {B{\'e}thermin}, {Blain},
  {Boselli}, {Bridge}, {Brisbin}, {Buat}, {Burgarella},
  {Castro-Rodr{\'{\i}}guez}, {Cava}, {Chanial}, {Cirasuolo}, {Clements},
  {Conley}, {Conversi}, {Cooray}, {Dowell}, {Dubois}, {Dwek}, {Dye}, {Eales},
  {Elbaz}, {Farrah}, {Feltre}, {Ferrero}, {Fiolet}, {Fox}, {Franceschini},
  {Gear}, {Giovannoli}, {Glenn}, {Gong}, {Gonz{\'a}lez Solares}, {Griffin},
  {Halpern}, {Harwit}, {Hatziminaoglou}, {Heinis}, {Hurley}, {Hwang}, {Hyde},
  {Ibar}, {Ilbert}, {Isaak}, {Ivison}, {Lagache}, {Le Floc'h}, {Levenson},
  {Faro}, {Lu}, {Madden}, {Maffei}, {Magdis}, {Mainetti}, {Marchetti},
  {Marsden}, {Marshall}, {Mortier}, {Nguyen}, {O'Halloran}, {Omont}, {Page},
  {Panuzzo}, {Papageorgiou}, {Patel}, {Pearson}, {P{\'e}rez-Fournon}, {Pohlen},
  {Rawlings}, {Raymond}, {Rigopoulou}, {Riguccini}, {Rizzo}, {Rodighiero},
  {Roseboom}, {Rowan-Robinson}, {S{\'a}nchez Portal}, {Schulz}, {Scott},
  {Seymour}, {Shupe}, {Smith}, {Stevens}, {Symeonidis}, {Trichas}, {Tugwell},
  {Vaccari}, {Valtchanov}, {Vieira}, {Viero}, {Vigroux}, {Wang}, {Ward},
  {Wardlow}, {Wright}, {Xu}, \& {Zemcov}}]{Oliver:Hermes}
{Oliver}, S.~J., {Bock}, J., {Altieri}, B., {et~al.} 2012, \mnras, 424, 1614

\bibitem[{{Ott}(2010)}]{Ott:2010}
{Ott}, S. 2010, in Astronomical Society of the Pacific Conference Series, Vol.
  434, Astronomical Data Analysis Software and Systems XIX, ed. Y.~{Mizumoto},
  K.-I. {Morita}, \& M.~{Ohishi}, 139

\bibitem[{{Park} \& {Choi}(2009)}]{Park:2009}
{Park}, C., \& {Choi}, Y.-Y. 2009, \apj, 691, 1828

\bibitem[{{Park} {et~al.}(2008){Park}, {Gott}, \& {Choi}}]{Park:2008}
{Park}, C., {Gott}, III, J.~R., \& {Choi}, Y.-Y. 2008, \apj, 674, 784

\bibitem[{{Pilbratt} {et~al.}(2010){Pilbratt}, {Riedinger}, {Passvogel},
  {Crone}, {Doyle}, {Gageur}, {Heras}, {Jewell}, {Metcalfe}, {Ott}, \&
  {Schmidt}}]{Herschel}
{Pilbratt}, G.~L., {Riedinger}, J.~R., {Passvogel}, T., {et~al.} 2010, \aap,
  518, L1

\bibitem[{{Poglitsch} {et~al.}(2010){Poglitsch}, {Waelkens}, {Geis},
  {Feuchtgruber}, {Vandenbussche}, {Rodriguez}, {Krause}, {Renotte}, {van
  Hoof}, {Saraceno}, {Cepa}, {Kerschbaum}, {Agn{\`e}se}, {Ali}, {Altieri},
  {Andreani}, {Augueres}, {Balog}, {Barl}, {Bauer}, {Belbachir}, {Benedettini},
  {Billot}, {Boulade}, {Bischof}, {Blommaert}, {Callut}, {Cara}, {Cerulli},
  {Cesarsky}, {Contursi}, {Creten}, {De Meester}, {Doublier}, {Doumayrou},
  {Duband}, {Exter}, {Genzel}, {Gillis}, {Gr{\"o}zinger}, {Henning},
  {Herreros}, {Huygen}, {Inguscio}, {Jakob}, {Jamar}, {Jean}, {de Jong},
  {Katterloher}, {Kiss}, {Klaas}, {Lemke}, {Lutz}, {Madden}, {Marquet},
  {Martignac}, {Mazy}, {Merken}, {Montfort}, {Morbidelli}, {M{\"u}ller},
  {Nielbock}, {Okumura}, {Orfei}, {Ottensamer}, {Pezzuto}, {Popesso},
  {Putzeys}, {Regibo}, {Reveret}, {Royer}, {Sauvage}, {Schreiber}, {Stegmaier},
  {Schmitt}, {Schubert}, {Sturm}, {Thiel}, {Tofani}, {Vavrek}, {Wetzstein},
  {Wieprecht}, \& {Wiezorrek}}]{PACS:2010}
{Poglitsch}, A., {Waelkens}, C., {Geis}, N., {et~al.} 2010, \aap, 518, L2

\bibitem[{{Renaud} {et~al.}(2014){Renaud}, {Bournaud}, {Kraljic}, \&
  {Duc}}]{Renaud:2014}
{Renaud}, F., {Bournaud}, F., {Kraljic}, K., \& {Duc}, P.-A. 2014, \mnras, 442,
  L33

\bibitem[{{Rieke} {et~al.}(2004){Rieke}, {Young}, {Engelbracht}, {Kelly},
  {Low}, {Haller}, {Beeman}, {Gordon}, {Stansberry}, {Misselt}, {Cadien},
  {Morrison}, {Rivlis}, {Latter}, {Noriega-Crespo}, {Padgett}, {Stapelfeldt},
  {Hines}, {Egami}, {Muzerolle}, {Alonso-Herrero}, {Blaylock}, {Dole}, {Hinz},
  {Le Floc'h}, {Papovich}, {P{\'e}rez-Gonz{\'a}lez}, {Smith}, {Su}, {Bennett},
  {Frayer}, {Henderson}, {Lu}, {Masci}, {Pesenson}, {Rebull}, {Rho}, {Keene},
  {Stolovy}, {Wachter}, {Wheaton}, {Werner}, \& {Richards}}]{Rieke:2004}
{Rieke}, G.~H., {Young}, E.~T., {Engelbracht}, C.~W., {et~al.} 2004, \apjs,
  154, 25

\bibitem[{{Roche} {et~al.}(1991){Roche}, {Aitken}, {Smith}, \&
  {Ward}}]{Roche:AGN}
{Roche}, P.~F., {Aitken}, D.~K., {Smith}, C.~H., \& {Ward}, M.~J. 1991, \mnras,
  248, 606

\bibitem[{{Roehlly} {et~al.}(2014){Roehlly}, {Burgarella}, {Buat}, {Boquien},
  {Ciesla}, \& {Heinis}}]{PCIGALE}
{Roehlly}, Y., {Burgarella}, D., {Buat}, V., {et~al.} 2014, in Astronomical
  Society of the Pacific Conference Series, Vol. 485, Astronomical Data
  Analysis Software and Systems XXIII, ed. N.~{Manset} \& P.~{Forshay}, 347

\bibitem[{Roehlly {et~al.}(2012)Roehlly, Burgarella, Buat, Giovannoli, Noll,
  {et~al.}}]{Roehlly:2011de}
Roehlly, Y., Burgarella, D., Buat, V., {et~al.} 2012, ASP Conf.Ser., 461, 569

\bibitem[{{Scudder} {et~al.}(2012){Scudder}, {Ellison}, {Torrey}, {Patton}, \&
  {Mendel}}]{Scudder:2012}
{Scudder}, J.~M., {Ellison}, S.~L., {Torrey}, P., {Patton}, D.~R., \& {Mendel},
  J.~T. 2012, \mnras, 426, 549

\bibitem[{{Skrutskie} {et~al.}(2006){Skrutskie}, {Cutri}, {Stiening},
  {Weinberg}, {Schneider}, {Carpenter}, {Beichman}, {Capps}, {Chester},
  {Elias}, {Huchra}, {Liebert}, {Lonsdale}, {Monet}, {Price}, {Seitzer},
  {Jarrett}, {Kirkpatrick}, {Gizis}, {Howard}, {Evans}, {Fowler}, {Fullmer},
  {Hurt}, {Light}, {Kopan}, {Marsh}, {McCallon}, {Tam}, {Van Dyk}, \&
  {Wheelock}}]{2mass:mission}
{Skrutskie}, M.~F., {Cutri}, R.~M., {Stiening}, R., {et~al.} 2006, \aj, 131,
  1163

\bibitem[{Stern {et~al.}(2012)Stern, Assef, Benford, Blain, Cutri, Dey,
  Eisenhardt, Griffith, Jarrett, Lake, Masci, Petty, Stanford, Tsai, Wright,
  Yan, Harrison, \& Madsen}]{stern:2012}
Stern, D., Assef, R.~J., Benford, D.~J., {et~al.} 2012, The Astrophysical
  Journal, 753, 30

\bibitem[{{Toomre} \& {Toomre}(1972)}]{ToomreToomre:1972}
{Toomre}, A., \& {Toomre}, J. 1972, \apj, 178, 623

\bibitem[{{Traficante} {et~al.}(2011){Traficante}, {Calzoletti}, {Veneziani},
  {Ali}, {de Gasperis}, {di Giorgio}, {Faustini}, {Ikhenaode}, {Molinari},
  {Natoli}, {Pestalozzi}, {Pezzuto}, {Piacentini}, {Piazzo}, {Polenta}, \&
  {Schisano}}]{Traficante:2011}
{Traficante}, A., {Calzoletti}, L., {Veneziani}, M., {et~al.} 2011, \mnras,
  416, 2932

\bibitem[{{Wright} {et~al.}(2010){Wright}, {Eisenhardt}, {Mainzer}, {Ressler},
  {Cutri}, {Jarrett}, {Kirkpatrick}, {Padgett}, {McMillan}, {Skrutskie},
  {Stanford}, {Cohen}, {Walker}, {Mather}, {Leisawitz}, {Gautier}, {McLean},
  {Benford}, {Lonsdale}, {Blain}, {Mendez}, {Irace}, {Duval}, {Liu}, {Royer},
  {Heinrichsen}, {Howard}, {Shannon}, {Kendall}, {Walsh}, {Larsen}, {Cardon},
  {Schick}, {Schwalm}, {Abid}, {Fabinsky}, {Naes}, \& {Tsai}}]{Wise:mission}
{Wright}, E.~L., {Eisenhardt}, P.~R.~M., {Mainzer}, A.~K., {et~al.} 2010, \aj,
  140, 1868

\bibitem[{{Xu} \& {Helou}(1996)}]{XuHelou}
{Xu}, C., \& {Helou}, G. 1996, \apj, 456, 163

\bibitem[{{Xu} \& {Sulentic}(1991)}]{Xu:1991}
{Xu}, C., \& {Sulentic}, J.~W. 1991, \apj, 374, 407

\bibitem[{{Xu} {et~al.}(2010){Xu}, {Domingue}, {Cheng}, {Lu}, {Huang}, {Gao},
  {Mazzarella}, {Cutri}, {Sun}, \& {Surace}}]{Xu:2010kpairs}
{Xu}, C.~K., {Domingue}, D., {Cheng}, Y.-W., {et~al.} 2010, \apj, 713, 330

\bibitem[{Xu {et~al.}(2012)Xu, Shupe, B{\'e}thermin, Aussel, Berta, Bock,
  Bridge, Conley, Cooray, Elbaz, Franceschini, Floc'h, Lu, Lutz, Magnelli,
  Marsden, Oliver, Pozzi, Riguccini, Schulz, Scoville, Vaccari, Vieira, Wang,
  \& Zemcov}]{XuShupe:2012}
Xu, C.~K., Shupe, D.~L., B{\'e}thermin, M., {et~al.} 2012, The Astrophysical
  Journal, 760, 72

\bibitem[{{York} {et~al.}(2000){York}, {Adelman}, {Anderson}, {Anderson},
  {Annis}, {Bahcall}, {Bakken}, {Barkhouser}, {Bastian}, {Berman}, {Boroski},
  {Bracker}, {Briegel}, {Briggs}, {Brinkmann}, {Brunner}, {Burles}, {Carey},
  {Carr}, {Castander}, {Chen}, {Colestock}, {Connolly}, {Crocker}, {Csabai},
  {Czarapata}, {Davis}, {Doi}, {Dombeck}, {Eisenstein}, {Ellman}, {Elms},
  {Evans}, {Fan}, {Federwitz}, {Fiscelli}, {Friedman}, {Frieman}, {Fukugita},
  {Gillespie}, {Gunn}, {Gurbani}, {de Haas}, {Haldeman}, {Harris}, {Hayes},
  {Heckman}, {Hennessy}, {Hindsley}, {Holm}, {Holmgren}, {Huang}, {Hull},
  {Husby}, {Ichikawa}, {Ichikawa}, {Ivezi{\'c}}, {Kent}, {Kim}, {Kinney},
  {Klaene}, {Kleinman}, {Kleinman}, {Knapp}, {Korienek}, {Kron}, {Kunszt},
  {Lamb}, {Lee}, {Leger}, {Limmongkol}, {Lindenmeyer}, {Long}, {Loomis},
  {Loveday}, {Lucinio}, {Lupton}, {MacKinnon}, {Mannery}, {Mantsch}, {Margon},
  {McGehee}, {McKay}, {Meiksin}, {Merelli}, {Monet}, {Munn}, {Narayanan},
  {Nash}, {Neilsen}, {Neswold}, {Newberg}, {Nichol}, {Nicinski}, {Nonino},
  {Okada}, {Okamura}, {Ostriker}, {Owen}, {Pauls}, {Peoples}, {Peterson},
  {Petravick}, {Pier}, {Pope}, {Pordes}, {Prosapio}, {Rechenmacher}, {Quinn},
  {Richards}, {Richmond}, {Rivetta}, {Rockosi}, {Ruthmansdorfer}, {Sandford},
  {Schlegel}, {Schneider}, {Sekiguchi}, {Sergey}, {Shimasaku}, {Siegmund},
  {Smee}, {Smith}, {Snedden}, {Stone}, {Stoughton}, {Strauss}, {Stubbs},
  {SubbaRao}, {Szalay}, {Szapudi}, {Szokoly}, {Thakar}, {Tremonti}, {Tucker},
  {Uomoto}, {Vanden Berk}, {Vogeley}, {Waddell}, {Wang}, {Watanabe},
  {Weinberg}, {Yanny}, {Yasuda}, \& {SDSS Collaboration}}]{SDSS}
{York}, D.~G., {Adelman}, J., {Anderson}, Jr., J.~E., {et~al.} 2000, \aj, 120,
  1579

\end{thebibliography}

\clearpage

 %%%%%%%%%%%%%%%%%%%%%%%%%%%%%%%%%%%%%%%%%%%%%%%%%
 	\clearpage

 \end{document}